\documentclass[twoside,twocolumn,9pt]{article}

\usepackage{extsizes}
\usepackage[left=1.5cm, right=1.5cm, top=1.785cm, bottom=2.0cm]{geometry}
\usepackage{balance}
\usepackage[utf8]{inputenc} % allow utf-8 input
\usepackage{hyperref}       % hyperlinks
\usepackage{url}            % simple URL typesetting
\usepackage{booktabs}       % professional-quality tables
\usepackage{amsfonts}       % blackboard math symbols
\usepackage{nicefrac}       % compact symbols for 1/2, etc.
\usepackage{microtype}      % microtypography
\usepackage{droidsans}
\usepackage{graphicx}
\usepackage{amsmath}
\usepackage{bbm}
\usepackage{mathtools}
\usepackage{authblk}
\usepackage{xcolor}

\usepackage[super,sort&compress,comma]{natbib}

\title{Topological Data Analysis of Collective and Individual Epithelial Cells using Persistent Homology of Loops}

\author[1,2]{Dhananjay Bhaskar}
\author[3]{William Y. Zhang}
\author[1,2]{Ian Y. Wong\thanks{ian\_wong@brown.edu}}
\affil[1]{School of Engineering, Center for Biomedical Engineering, Brown University, 184 Hope St Box D, Providence, RI 02912}
\affil[2]{Data Science Initiative, Brown University, 184 Hope St Box D, Providence, RI 02912}
\affil[3]{Department of Computer Science, Brown University, 184 Hope St Box D, Providence, RI 02912}

\begin{document}

\twocolumn[
  \begin{@twocolumnfalse}

\maketitle

\begin{abstract}
Interacting, self-propelled particles such as epithelial cells can dynamically self-organize into complex multicellular patterns, which are challenging to classify without \emph{a priori} information. Classically, different phases and phase transitions have been described based on local ordering, which may not capture structural features at larger length scales. Instead, topological data analysis (TDA) determines the stability of spatial connectivity at varying length scales (i.e. persistent homology), and can compare different particle configurations based on the ``cost'' of reorganizing one configuration into another. Here, we demonstrate a topology-based machine learning approach for unsupervised profiling of individual and collective phases based on large-scale loops. We show that these topological loops (i.e. dimension 1 homology) are robust to variations in particle number and density, particularly in comparison to connected components (i.e. dimension 0 homology). We use TDA to map out phase diagrams for simulated particles with varying adhesion and propulsion, at constant population size as well as when proliferation is permitted. Next, we use this approach to profile our recent experiments on the clustering of epithelial cells in varying growth factor conditions, which are compared to our simulations. Finally, we characterize the robustness of this approach at varying length scales, with sparse sampling, and over time. Overall, we envision TDA will be broadly applicable as a model-agnostic approach to analyze active systems with varying population size, from cytoskeletal motors to motile cells to flocking or swarming animals.
\end{abstract}

\end{@twocolumnfalse} \vspace{0.6cm}
]

\section{Introduction}
Collective behaviors emerge from multi-particle interactions, resulting in rich self-organizing patterns.\cite{Vicsek:2012gp} For instance, epithelial cells assemble into tightly connected multicellular layers  due to strong cell-cell and cell-matrix adhesions, representing a fascinating system of non-equilibrium dynamics.\cite{Xi:2018fh} Moreover, multicellular clusters can ``scatter'' as migratory individuals in response to biochemical stimuli,\cite{deRooij:2005gc,Loerke:2012jz,Maruthamuthu:2014dq} analogous to an epithelial-mesenchymal transition.\cite{Wong:2014jma} Instead, dispersed and motile individuals can transition towards collective migration and ultimately arrested states, analogous to a ``jamming'' transition.\cite{Szabo:2006gr,Angelini:2011dm,Mehes:2012fy,Suaris:2013ei,Park:2015ih,Bi:2015hh,Garcia:2015be,Bi:2016bs,GamboaCastro:2016da,Duclos:2016jm,Atia:2018bp, Leggett:2019, KIM2020706}

Epithelial cells can be computationally modeled as self-propelled particles in two dimensional space.\cite{Camley:2017io} Experimentally, individual epithelial cells exhibit persistent migration alternating with random re-orientation to a new direction,\cite{Potdar:2009ex} analogous to ``run-and-tumble'' behaviors observed in bacteria.\cite{Tailleur:2008kd} In comparison, active Brownian particles also exhibit persistent motion, but their re-orientation is governed by rotational Brownian motion,\cite{Fodor:2018en} which can yield similar trends in the limit of large timescales.\cite{Cates:2013ia} At higher densities, cells have been treated as disks with some isotropic repulsive potential \cite{PhysRevLett.100.248702,Garcia:2015be,PhysRevE.84.040301,10.1371/journal.pcbi.1002944,10.1371/journal.pcbi.1004670,PhysRevE.91.032706,Szabo:2006gr,Yeo:2015jz,Volkening:2015dj,Camley:2016gy,Camley:2017gk,PhysRevLett.118.158105,C6SM02580C,george2017connecting,McCusker_2019, McGuirl201917763}, which can further interact via attractive potentials or local alignment (e.g. Vicsek model),\cite{Szabo:2006gr} resulting in spatiotemporal correlations in position and velocity. However, a potentially confounding behavior of living systems is that the size of the population gradually changes due to proliferation or death,\cite{C6SM02580C,Tjhung:2020dy} which can complicate quantitative comparisons across different conditions over time.

Topological data analysis (TDA) is an emerging mathematical framework for visualizing the underlying ``structure'' of high-dimensional datasets based on the
spatial connectivity between discrete points.\cite{Carlsson:2020ko} TDA determines the robustness of connectivity between points over a range of spatial scales (i.e. persistent homology), which are represented by pairwise connected components (dimension 0 homology), connected loops around an empty area (dimension 1 homology), etc., and are summarized by a persistence diagram or barcode.\cite{Amezquita:2020bi} The topological similarity between spatial configurations can be determined by comparing their respective persistence diagrams, and the ``cost'' of rearranging one diagram to resemble another.\cite{herbertedelsbrunner2009} Topological approaches are gaining interest to visualize swarming or patterning behaviors in living entities,\cite{Topaz:2015gd,Ulmer:2019hx,Bhaskar:2019hj,McGuirl201917763} as well as percolation thresholds in 2D disk packing.\cite{Speidel:2018gn} Previous approaches have largely focused on counting connected components (dimension 0 homology) for populations with constant size that are space-filling (confluent).\cite{atienza2019persistent,skinner2020topological} However, the number of connected components will change for varying population size, which requires some arbitrary normalization in order to meaningfully compare different persistence diagrams \cite{atienza2019persistent}, or be weighted by population number and local density.\cite{skinner2020topological} Thus, it remains challenging to implement unbiased and unsupervised analysis of (dis)ordered and collective phases in active matter systems with varying population size and density.

Here, we show that TDA enables unbiased and unsupervised classification of collective and individual phases in epithelial cells based on simulated and experimental data. Our major innovation is to use topologically connected loops to summarize multicellular architectures over larger length scales, which encodes additional information that is not captured by more localized metrics based on order parameters or topologically connected components. We first applied this approach on a training set of interacting self-propelled particles with varying adhesion at constant population size. We subsequently generalize for interacting self-propelled particles that exhibit significant proliferation over the course of the simulation. We show that TDA can be utilized for experimental data based on tracking epithelial cell nuclei, accurately classified experimental results from different biochemical treatments, and mapping back to the most topologically similar simulations. Finally, we investigate the effectiveness of TDA by measuring topological differences when sampling at varying length scales, sparse data, and over time. Since this approach is robust against biological variability such as population size and density, we anticipate that TDA will be broadly applicable for visualizing how living units migrate, proliferate, and interact across length scales from molecular motors to mammalian cells to animals.

\section{Topological Data Analysis of Particle Configurations based on Persistent Loops}

\begin{figure}[h!]
\centering
\includegraphics[width=9cm]{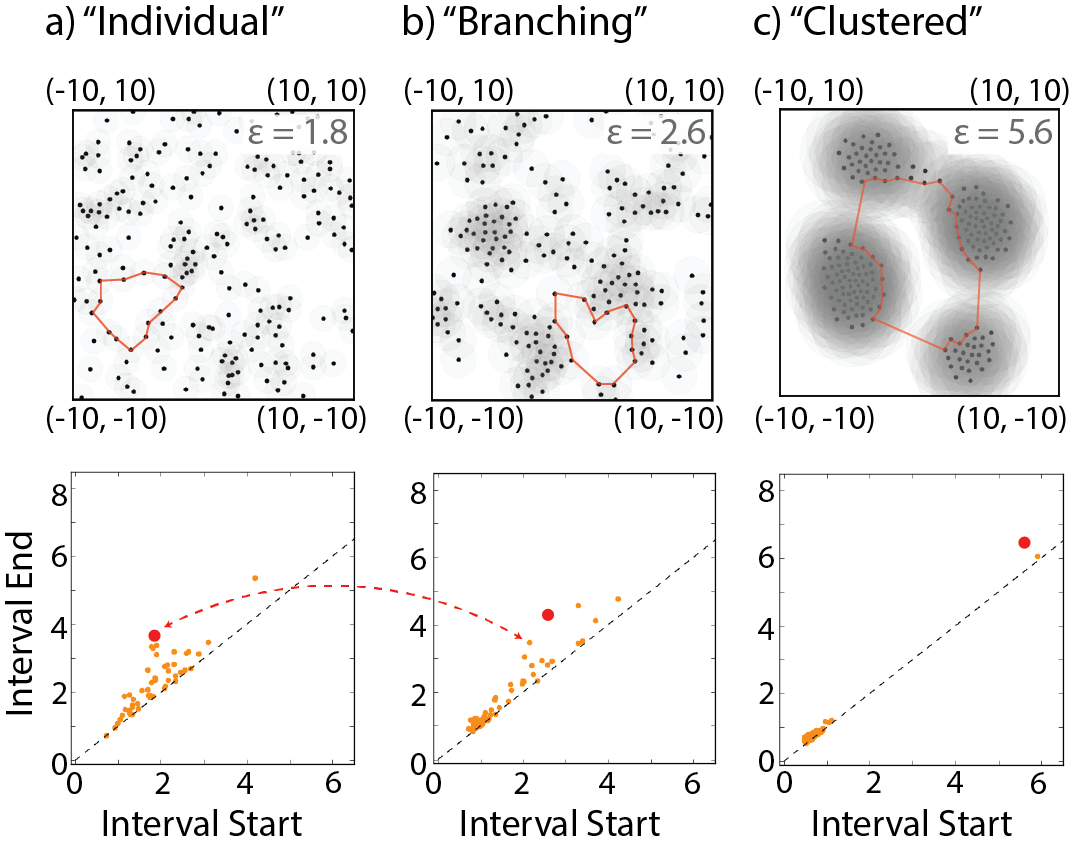}
\caption{\textbf{Comparison of representative particle configurations and corresponding persistence diagrams}. a) Representative individual particle configuration. Gray shading shows particle connectivity at $\epsilon = 1.8$. b) Representative branching network particle configuration. Gray shading shows particle connectivity at $\epsilon = 2.6$. c) Representative clustered particle configuration. Gray shading shows particle connectivity at $\epsilon =  5.6$. Red loops in the particle configuration correspond to persistent features highlighted in red in the persistence diagram. Dashed red arrow indicates matching of persistent features for Wasserstein distance computation.}
\label{fig:newfig1}\vspace{-0.1in}
\end{figure}

Our approach sought to classify different spatial configurations of particle centroids based on the presence and persistence of topological loops. For instance, two points that lie within some cutoff distance $\epsilon$ can be linked together by an edge (forming a connected component characterized by Betti number, $\beta_0=1$). Moreover, a circular set of points that are pairwise within a separation distance $\epsilon$ can be linked into a closed loop enclosing a one-dimensional hole (characterized by Betti number, $\beta_1=1$). As an illustrative example, we consider three representative particle configurations corresponding to individually dispersed, branching network and compact clusters. A closed loop surrounding an empty region, which persists across a wide range of cutoff distances $\epsilon$ is highlighted in red for each configuration (\textbf{Fig.~\ref{fig:newfig1}}). Persistence diagrams are then used to visualize the cutoff distances $\epsilon$ for which a topological loop appears or disappears. For instance, the loop shown in \textbf{Fig.~\ref{fig:newfig1}}a first appears at $\epsilon = 1.8$ ($x$-coordinate) and disappears at $\epsilon = 3.7$ ($y$-coordinate). There exist many additional topological loops that are less persistent, since they appear and then disappear for smaller differences of $\epsilon$, and are thus represented by ``noisy'' points of less interest close to the diagonal. The construction of persistence diagrams and the analogous topological barcodes is reviewed in greater detail in Supplementary Information (\textbf{Note S1 and Fig.~\ref{fig:barcodediagram}}).

Interacting, self-propelled particles can self-organize into a variety of spatial configurations based on the relative strengths of their inter-particle interactions and propulsion, which will be defined quantitatively in the next section. For instance, particles with comparable propulsion and adhesion forces form branched networks that are more space filling. Small empty regions exist between these branching structures, which can be enclosed by topological loops with intermediate diameter (mean value $\approx 2.5$) (\textbf{Fig.~\ref{fig:newfig1}b}). In comparison, particles with stronger adhesion and weaker propulsion aggregate as compact clusters that are more separated (\textbf{Fig.~\ref{fig:newfig1}c}). Larger empty regions exist between these clusters, which are enclosed by topological loops of larger diameter (mean value $\approx 5.4$). The ``similarity'' between persistence diagrams can be computed based on a Wasserstein distance, which measures the ``cost'' of rearranging features on one diagram to resemble another. Further details of the Wasserstein distance calculation are described in Supplementary Information (\textbf{Note S1 and Fig.~\ref{fig:SI_Fig2}}).

Computationally, the persistence of topological features was quantified by extracting a persistence diagram using the Vietoris-Rips complex, implemented in Julia's Eirene package.\cite{henselmanghristl6} Pairwise Wasserstein distances were then computed using Eirene to compare different particle configurations obtained from simulations or microscopy images. Classification was performed with complete-linkage hierarchical clustering using SciPy.\cite{2020SciPy-NMeth}

\section{Computational Model}
\begin{figure}
\centering
\includegraphics[width=7.4cm]{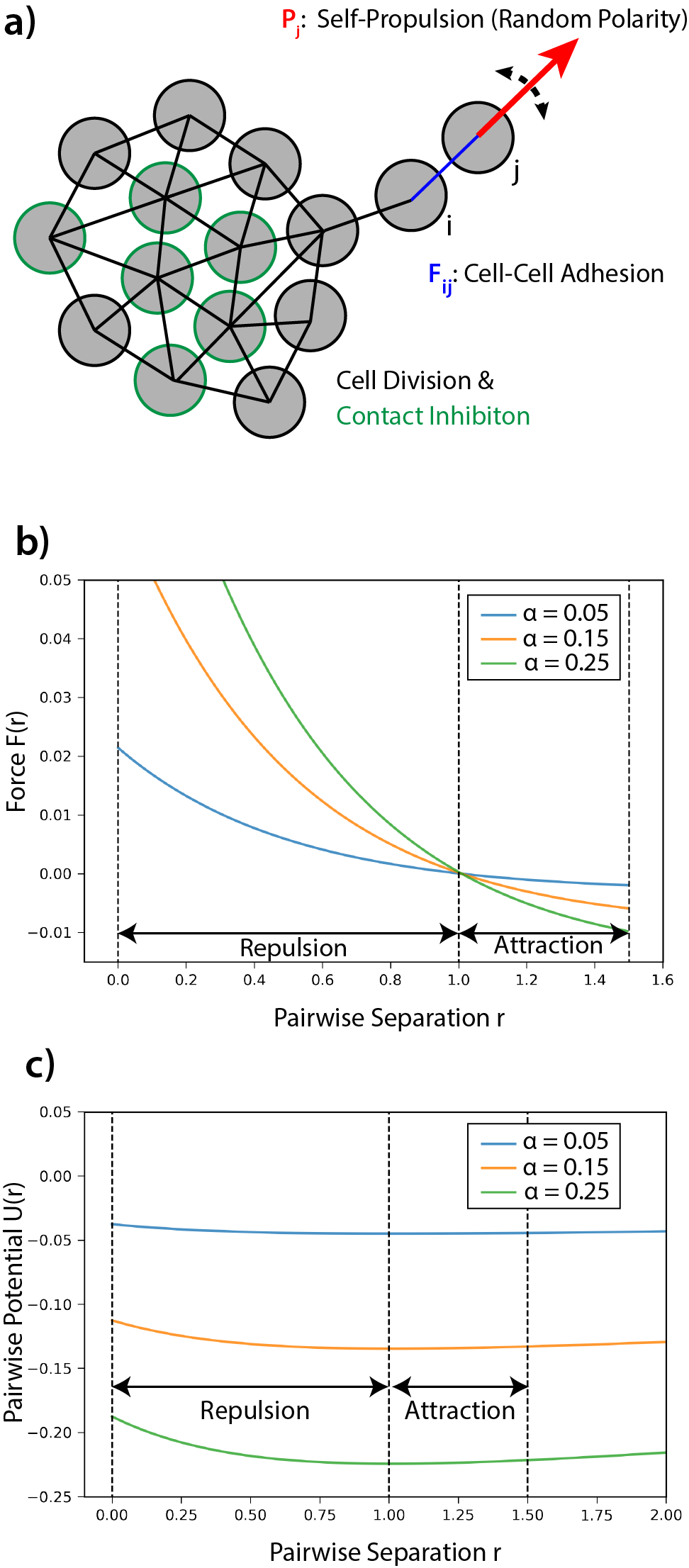}
\caption{\textbf{Self-propelled particle model.} (a) Cells were represented as disks propelled at constant force $\mathbf{P}$ (with random orientation) and cell-cell adhesion force $\mathbf{F}$. A ``bond'' was drawn between two cells if they were within distance $r = 1$ of one another. Cells with 4 or more neighbors (outlined in green) were not permitted to proliferate. (b) The adhesion force exerted on cell $i$ (located at $r = 0$) due to neighboring cell $j$, $\mathbf{F}_{ij}$, as a function of radial distance $r$ was plotted for various values of the adhesion parameter $\alpha$. Long-range attractive force (pointing inwards towards cell $i$) was negative and short-range repulsion (pointing outwards, away from cell $i$) was positive. Note that the attraction force was cut-off at $r=r_{\text{max}}=1.5$. (c) The attraction-repulsion kernel $U$, was plotted for various adhesion parameter values, $\alpha$. Between $0 \le r \le r_{\text{max}}$, the kernel was minimized at $r=r_{\text{eq}}\approx1$, the equilibrium distance we use to define neighboring cells indicated by a ``bond" drawn between them.}
\label{fig:Figure1}\vspace{-0.2in}
\end{figure}

Our model represented epithelial cells as self-propelled particles with three features. First, particles travel at constant velocity but randomly polarized in new directions at constant intervals (offset to different times, i.e. run and tumble model),\cite{Fodor:2018en} consistent with experimental measurements.\cite{Potdar:2009ex} Second, particles interact with nearby neighbors through a short-range repulsion corresponding to the particle radius, as well as a tunable attractive interaction. Third, particles can proliferate at regular intervals (offset to different starting times), unless surrounded by four or more neighbors (i.e. contact inhibition of proliferation)  (\textbf{Fig.~\ref{fig:Figure1}a}). The following model parameters were calibrated against our previous experimental measurements \cite{Leggett:2019}, although the model presented here uses a different interparticle interactions.

Simulations were initialized with $200$ particles randomly placed on a square domain ($[-10,10]\times[-10,10]$) with periodic boundary conditions. To ensure that particles were not too close together at initialization, a rejection sampling algorithm was used. At least three simulations with identical parameter values but distinct initial conditions were run over $200,000$ timesteps using the following over-damped equation of motion:

\begin{equation}
\mathbf{r}_{i}^{t+\Delta t} = \mathbf{r}_{i}^{t} + \frac{\Delta t}{\eta} \left( \mathbf{P}_{i}^{t} + \sum_{\mathclap{\substack{j=1\\j\neq i}}}^{N(t)} \mathbf{F}_{ij}^{t} \right)
\end{equation}
where $\mathbf{r}_{i}^{t}$ denoted the position vector of particle $i$ at time $t$, $\Delta t$ denoted the time step (default value $\Delta t = 0.02$), $\eta$ represented a drag coefficient (with default value $\eta = 1$) and $N(t)$ was the number of particles at time $t$.

The second term $\mathbf{P}_{i}^{t}$ represented a self-propulsion force acting on particle $i$ at time $t$, with constant magnitude $P$ varying from $0.009 - 0.025$. The direction (``polarization'') $\hat{\theta}_i^t$ of this self-propulsion force   was chosen uniformly at random once every $\tau_p = 2,500$ timesteps. To prevent cells from repolarizing at the same time, an offset (a random value chosen uniformly between $0$ and $500$ timesteps) was initially subtracted from the total time to repolarization for each cell.

The third term $\mathbf{F}_{ij}$ represented pairwise cell-cell interactions for cell $i$ with other cells $j$, and is plotted in \textbf{Fig.~\ref{fig:Figure1}b,c} for three representative adhesion values $\alpha = 0.05, 0.15, 0.25$:

\begin{equation}
\mathbf{F}_{ij} = -\nabla U(\|\mathbf{r}_{j}-\mathbf{r}_{i}\|)\frac{\mathbf{r}_{j}-\mathbf{r}_{i}}{\|\mathbf{r}_{j}-\mathbf{r}_{i}\|}  \mathbbm{1}_{0 \le \| \mathbf{r}_{j} - \mathbf{r}_{i} \| \le r_{\text{max}}}
\end{equation}

where the attraction-repulsion kernel $U$ governed the overall magnitude of adhesion and repulsion between any pair of cells. Note that the cell-cell interaction is only active at radial distances between $0$ and $r_{\text{max}} = 1.5$, preventing a given cell from attracting distant cells beyond 2 cell radii. The force of this interaction was obtained by computing the gradient of this potential function, which included four parameters:

\begin{equation}
U(r) = -c_A e^{-r/L_A} + c_R e^{-r/L_R}
\end{equation}

which specified length scales for long range attraction ($L_A=14.0$) and short range repulsion ($L_R=0.5$) as well as the relative strength of attraction and repulsion ($c_A=\alpha$ and $c_R=0.25\alpha$, respectively). This potential has a minimum at $r_{eq} \approx 1$. The parameter $\alpha$, varying from $0.05 - 0.25$, controls the strength of the adhesion and repulsion force. The system can be parameterized using two nondimensional variables: the Peclet number $Pe = P\tau_p/\eta r_{eq}$, and a dimensionless adhesion (scaled by self-propulsion force) $A = \Delta U / P \Delta r$, where $\Delta U$ is the energy cost to move a particle from $r_{eq}$ to $r_{max}$, which we define as $\Delta r$. Further details of this calculation are provided in Supplementary Information (\textbf{Note S2}, \textbf{Fig.~\ref{fig:SI_FBD}}).

For some simulations, proliferation was also included by adding a new ``daughter'' particle placed close to the ``parent'' with a polarization vector in the opposite direction. For all particles, the total cell cycle time was the same ($50,000$ timesteps), with an initial randomly chosen offset (between $0$ and $10,000$ timesteps) to avoid biologically unrealistic synchrony in cell division. Particles with $4$ or more nearest neighbors were not permitted to undergo division, representing contact inhibition of proliferation.\cite{McClatchey:2012in} Particles were defined as neighbors if they were positioned within a distance of $r = 1$ from each other, which is indicated by plotting a ``bond'' between these particles. A group of $4$ or more neighboring cells, with cell-cell adhesion bonds that persist over many simulation time-steps was considered as a cluster.

All simulations were conducted at the Brown Center for Computation and Visualization.  Both simulation code and TDA code is open-sourced on GitHub (contact authors for access).

\section{Results}

\subsection{Classifying Individual and Clustered Phases Based on Topological Loops at Constant Population Size}
\begin{figure}
\centering
\includegraphics[width=7.7cm]{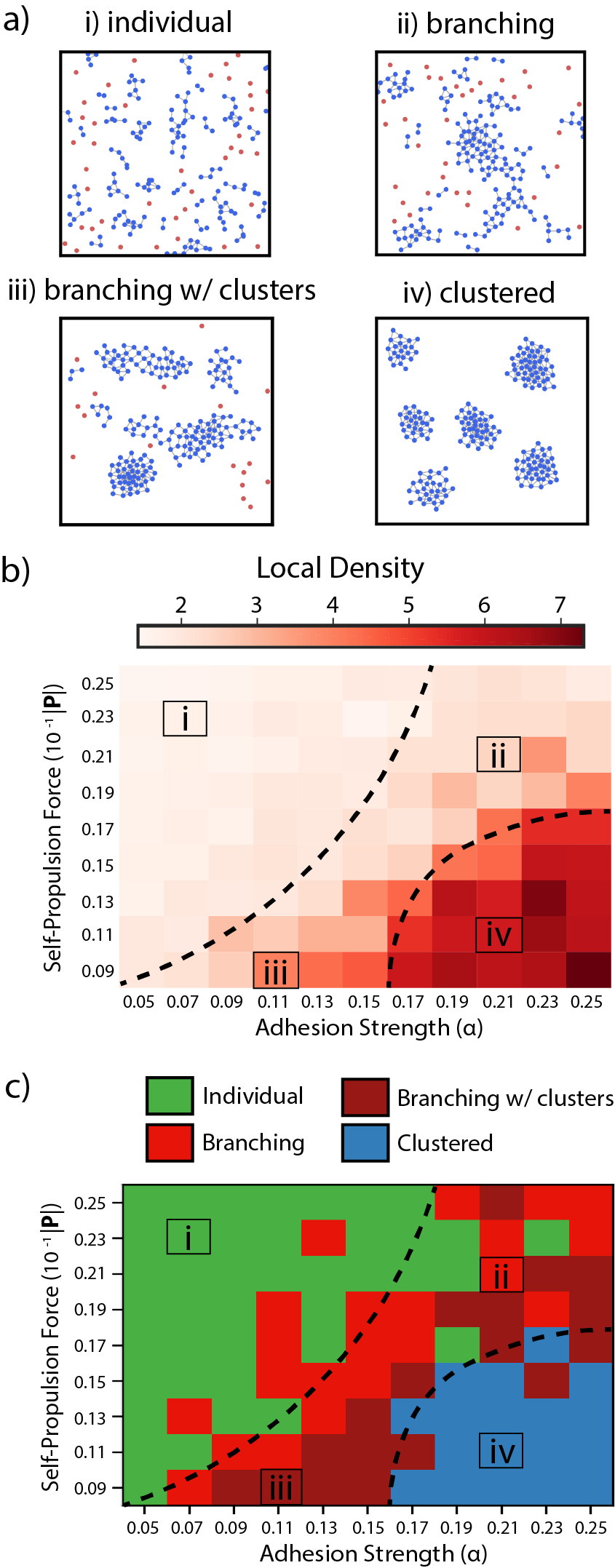}
\caption{\textbf{Individual and clustered phases exhibit distinct topological structure at constant population size.} (a) Snapshots of final configurations observed in simulations of the self-propelled particles for various adhesion and self-propulsion values. (b) Comparison of individual, branching, and clustered phases based on counting the ensemble averaged number of nearest neighbors within $r = r_\text{eq}$. (c) Comparison of individual, branching, and clustered phases classified by TDA of persistent loops.}
\label{fig:Figure3}\vspace{-0.1in}
\end{figure}

First, we considered a system consisting of a fixed number of self-propelled particles, where the speed of the $i$th particle at time $t$ was specified by a self-propulsion force $\mathbf{P}_i^t$ with random orientation, which repolarized in a different direction after some duration (i.e. run and tumble model). We further varied the relative adhesive interactions through the parameter $\alpha$, which sets the magnitude of the pairwise potential. As the propulsion $P$ and adhesion strength $\alpha$ were varied, three representative phases were qualitatively observed at the completion of the simulation ($t = 4000 = 200,000\Delta t$). First, for strong propulsion $P$ and weak adhesion strength $\alpha$, particles remained individually dispersed or interacted transiently as unstable clusters (\textbf{Fig.~\ref{fig:Figure3}a,i}). Next, when propulsion $P$ and adhesion strength $\alpha$ were comparably strong or weak, a branching phase was observed where clusters exhibited extended morphology (\textbf{Fig.~\ref{fig:Figure3}a,ii,iii}). Finally, for weak propulsion $P$ and strong adhesion strength $\alpha$, all particles were incorporated within larger rounded clusters (\textbf{Fig.~\ref{fig:Figure3}a,iv}). It should be noted that the particle dynamics at the completion of the simulations had reached some steady state, where particles either remained as individuals throughout the simulation (\textbf{Fig.~\ref{fig:SI_Fig3}a}), were associated with a branching configuration in dynamic equilibrium (\textbf{Fig.~\ref{fig:SI_Fig3}b,c}) or isolated clusters (\textbf{Fig.~\ref{fig:SI_Fig3}d}), before the completion of the simulation.

These distinct phases were classified using an order parameter that counted the number of nearest neighbors within a distance of $r_{\text{eq}}$, representing the local particle density (\textbf{Fig.~\ref{fig:Figure3}b}). In the limit of strong self-propulsion $0.015 < P$ and weak adhesion strength $\alpha < 0.15$, particles were observed to be mostly migratory individuals (\textbf{Fig.~\ref{fig:Figure3}b,i}), with an ensemble averaged number of nearest neighbors  $\langle n\rangle \approx 1$. Instead, in the limit of weak self-propulsion $P < 0.015$ and strong adhesion strength $0.15 < \alpha$, particles were typically organized into compact clusters (\textbf{Fig.~\ref{fig:Figure3}b,iv}), with $\langle n\rangle \approx 5$ nearest neighbors. Finally, when self-propulsion $P$ and adhesion strength $\alpha$ were comparable between these two regimes, a branching phase was observed with $\langle n\rangle \approx 3$ (\textbf{Fig.~\ref{fig:Figure3}b,ii,iii}). In order to determine the statistical distribution of $\langle n\rangle$, these values were calculated for 10 independent simulations with different initial particle configurations, but identical propulsion $P$ and  adhesion strength $\alpha$. For instance, individual simulations with $\alpha = 0.07, P = 0.021$ typically showed a mean of $\langle n\rangle$ = 1.6. Moreover, branching simulations with $\alpha = 0.09, P = 0.011$ showed $\langle n\rangle$ = 2.2, and clustered simulations with $\alpha = 0.23, P = 0.009$ showed $\langle n\rangle$= 6.4 (\textbf{Fig.~\ref{fig:SI_Fig4}a}). For comparison, the system was parameterized using Peclet number ($Pe$) and nondimensional adhesion ($A$). At $Pe \approx 1$, a transition from clusters to individuals occurred, based on the competition between (persistent) self-propulsion and reorientation. Moreover, at $A \approx 1$, a transition from individuals to clusters occurred, based on the competition between adhesion and self-propulsion. The branching phase occurred when $Pe \sim A$  (\textbf{Fig.~\ref{fig:SI_nondim_phase_diag}a,b}). One drawback of this approach was that $\langle n\rangle$ was defined based on \emph{a priori} information, since the expected interparticle spacing required knowledge of the pairwise interaction potential.

We observed that individual, clustered, and branching phases exhibited quantitative differences in the number and characteristic diameter of topological loops. Typically, individual phases exhibited $31-46$ loops of diameter $\approx 1.4-5.4$, branched phases exhibited $19-31$ loops of diameter $\approx 1.4-6.0$, and clustered phases exhibited $1-3$ loops of diameter $\approx 3.8-6.9$ (\textbf{Fig.~\ref{fig:SI_Fig4}b}). We then computed the pairwise Wasserstein distances between loops for persistence diagrams from all 121 simulations, with varying propulsion $P$ and adhesion $\alpha$. Hierarchical clustering of Wasserstein distance grouped simulations by clustered, individual, branching, and a mixed branching + clusters phase along the diagonal (\textbf{Fig.~\ref{fig:SI_Fig5}}). This analysis also revealed several noteworthy off-diagonal entries, indicating some similarity between clustered and ``branching with clusters'' phases, as well as individual and branching phases (\textbf{Fig.~\ref{fig:SI_Fig5}}). Based on this classification, distinct parameter regimes were mapped out corresponding to individual, branching, branching with clusters, as well as clustered phases. Indeed, these phases calculated using TDA show good agreement with the phases defined based on nearest neighbors $\langle n\rangle$ (\textbf{Fig.~\ref{fig:Figure3}c}). Occasionally, some simulations within the individual phase were misclassified as branching (e.g. $\alpha = 0.11, P = 0.015-0.019)$, while other simulations in the branching phase were misclassified as individual (e.g. $\alpha = 0.19, P = 0.017; \alpha = 0.23, P = 0.023$). These simulations often exhibited local densities that are slightly higher (or lower) than other simulations within their region, suggesting that TDA may be picking up subtle differences in particle configuration. Overall, TDA can be used for unsupervised classification of individual, branching, and clustered phases in snapshots of self-propelled particles, in excellent agreement with the phases and phase boundaries defined by a predefined order parameter.

\subsection{Branching and Clustered Phases Based on Topological Loops in Proliferating Populations}
Next, we considered a system consisting of proliferating self-propelled particles, where a parent particle divided after a fixed duration ($50,000$ timesteps), randomly offset. This proliferation was implemented by maintaining the parent particle with the same velocity and direction, but adding a second daughter particle (close to the parent) moving with equal velocity but opposite in direction to the parent. Moreover, the parent particle could not divide if it had more than four neighbors, which mimics the contact inhibition of proliferation of epithelial cells at high density.\cite{McClatchey:2012in} The self-propulsion $P$ and adhesion $\alpha$ were again systematically varied over the same range as in the previous simulations without proliferation. Simulations were initialized with 100 particles and analyzed at the final timestep after $200,000 \Delta t$ $(t = 4000)$.

\begin{figure}
\centering
\includegraphics[width=8cm]{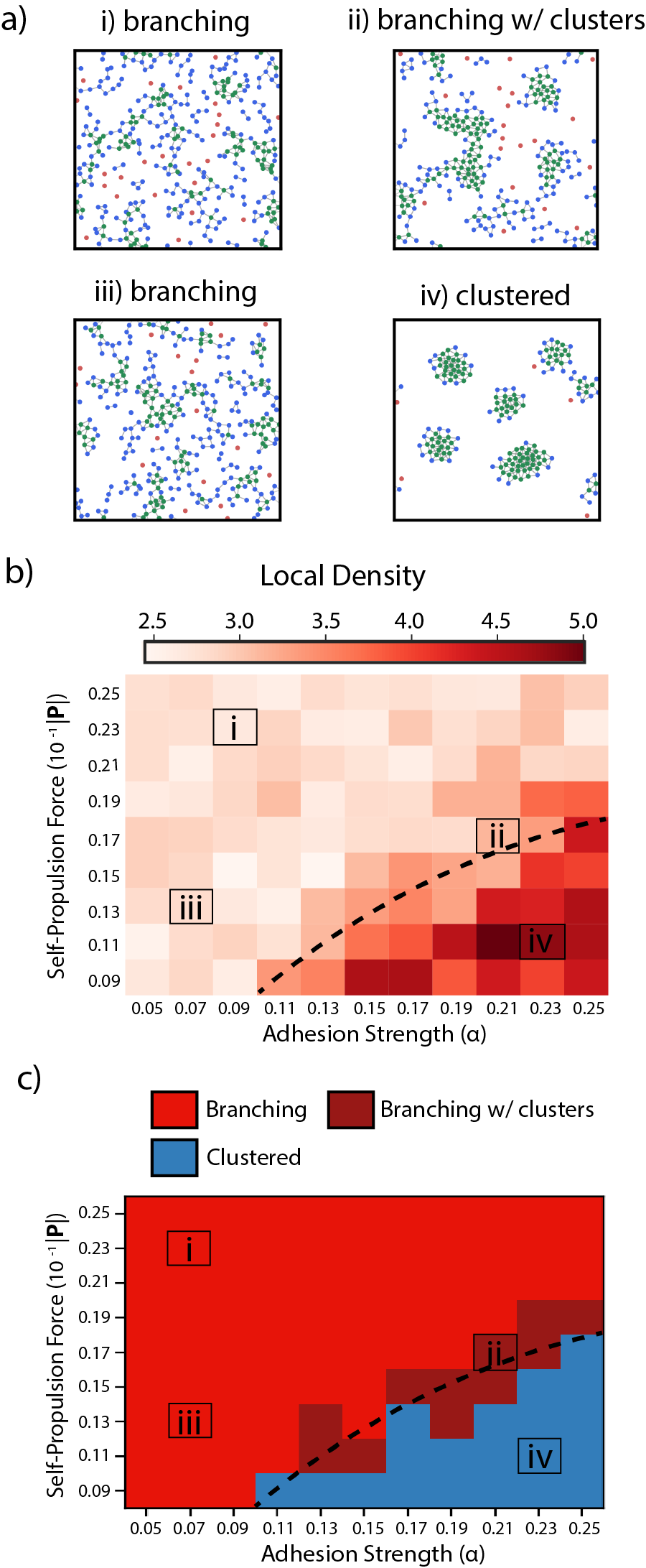}
\caption{\textbf{Branching and clustered phases exhibit distinct topological structure in proliferating populations.} (a) Snapshots of final configurations observed in simulations of the self-propelled particles for various adhesion and self-propulsion values. (b) Comparison of branching and clustered phases based on counting the ensemble averaged number of nearest neighbors within $r = r_\text{eq}$. (c) Comparison of branching and clustered phases classified by TDA of persistent loops.}
\label{fig:Figure4}\vspace{-0.1in}
\end{figure}

In the limit of weak adhesion $\alpha$, proliferating particles were observed as a branching phase with small clusters and $\langle n\rangle \approx 3$  (\textbf{Fig.~\ref{fig:Figure4}a,i,iii; b}), which differs from the migratory individuals observed previously without proliferation (\textbf{Fig.~\ref{fig:Figure3}a,i; b}). This difference occurred since individual cells (with few neighbors) were permitted to proliferate, whereas cells within a large cluster (with many neighbors) were not allowed to proliferate based on contact inhibition of proliferation. Next, when propulsion $P$ and adhesion $\alpha$ were comparable, a slightly different branching phase was observed with larger clusters that exhibited extended conformations and $\langle n\rangle \approx 2$ (\textbf{Fig.~\ref{fig:Figure4}a,ii; b}). Finally, for weak propulsion $P$ and strong adhesion $\alpha$, all particles were associated with clusters of compact morphology and $\langle n\rangle \approx 4$ (\textbf{Fig.~\ref{fig:Figure4}a,iv; b}). We further verified the variation in $\langle n \rangle$ by running simulations with different initial conditions but identical parameters for self-propulsion and adhesion (\textbf{Fig.~\ref{fig:SI_Fig4}c}). When this system is parameterized using nondimensional variables, these phase transitions again occurred at $Pe \approx 1$ and $A \approx 1$ (\textbf{Fig.~\ref{fig:SI_nondim_phase_diag}c,d}). Due to contact inhibition of proliferation, the particle dynamics and population size approached some steady state at the completion of the simulations, where particles either remained in a dynamic equilibrium within a branching network (\textbf{Fig.~\ref{fig:SI_Fig6}a,b}) or as isolated clusters (\textbf{Fig.~\ref{fig:SI_Fig6}c}), well before the completion of the simulation ($t = 4000 = 200,000\Delta t$). Nevertheless, it should be noted that population size varied from 160 - 360 particles across varying parameter values, with larger total numbers of particles at high self-propulsion and low adhesion, and decreasing particle numbers with decreasing self-propulsion and increasing adhesion, as more clusters formed. These relative differences in proliferation rate reduced the nearest neighbor counts for each phase (\textbf{Fig.~\ref{fig:SI_Fig6}c}), relative to nearest neighbor counts at constant population size (\textbf{Fig.~\ref{fig:SI_Fig6}a}).

We observed that clustered and branching phases in proliferating populations exhibited quantitative differences in the number and characteristic diameter of topological loops. Typically, branching simulations $(\alpha = 0.09, P = 0.009)$ exhibited $15-30$ loops of characteristic diameter $\approx 1.4-5.4$, while clustered simulations $(\alpha = 0.23, P = 0.007)$ exhibited $2-8$ loops of characteristic diameter $\approx 2.5-7.8$ (\textbf{Fig.~\ref{fig:SI_Fig4}d}). We again computed pairwise Wasserstein distances based on loops between persistence diagrams of all 121 simulations with varying self-propulsion $\mathbf{P}$ and adhesion $\alpha$. Hierarchical clustering of Wasserstein distances grouped simulations by branching and clustered phases along the diagonal (\textbf{Fig.~\ref{fig:SI_Fig7}}). Interestingly, the branching group was further divided into two branching subgroups with a ``branching with clusters'' subgroup in between (\textbf{Fig.~\ref{fig:SI_Fig7}}). The ``branching with clusters'' subgroup showed high similarity with both branching and clustered groups, as expected (\textbf{Fig.~\ref{fig:SI_Fig7}}). Mapping these classifications back to the phase diagram showed good agreement with the phases defined by the nearest neighbor order parameter (\textbf{Fig.~\ref{fig:Figure4}c}). Indeed, the top left, top right, and bottom left regions were classified as branching, the bottom right was classified as clustered, and some transition region of ``branching with clusters'' classified between them. These results show for the first time that TDA can perform unsupervised classification based on the presence of loops when population size varies significantly, showing quantitatively similar results as branching and clustered phases defined by some predetermined order parameter.

\subsection{Classifying Experimentally Measured Epithelial Cells after Varying Biochemical Treatments}
As a case study, we sought to classify our recent experimental measurements of mammary epithelial cells (MCF-10A) that transition from individuals to clusters when cultured in ``assay'' media with reduced concentrations of epidermal growth factor (EGF, 0.075 ng/mL).\cite{Leggett:2019} We previously showed that these cells exhibited slower proliferation and migration over 60 h, organizing over time into clusters with extended branching (fractal-like) architectures, analogous to diffusion-limited aggregation of non-living colloidal particles. These branching conformations were more pronounced after treatment with 4-hydroxytamoxifen (OHT), which activated EMT through an inducible Snail-estrogen receptor construct to drive leader cell formation,\cite{Wong:2014jma} relative to a DMSO control with more morphologically compact clusters. In comparison, cells cultured in ``growth'' media with considerably higher concentrations of EGF (20 ng/mL) remained highly migratory as individuals, before eventually proliferating over 60 h to fill the field of view as a confluent monolayer. Cells were localized using the the centroid of fluorescent nuclei (i.e. mCherry-H2B), which were detected as described previously.\cite{Leggett:2019} Persistence homology and pairwise Wasserstein distances were computed using the same methodology described above for analyzing simulation data. In combination with varying initial cell densities, these experimental measurements represent a more challenging test set for TDA-based classification.

\begin{figure*}
\centering
\includegraphics[width=17.1cm]{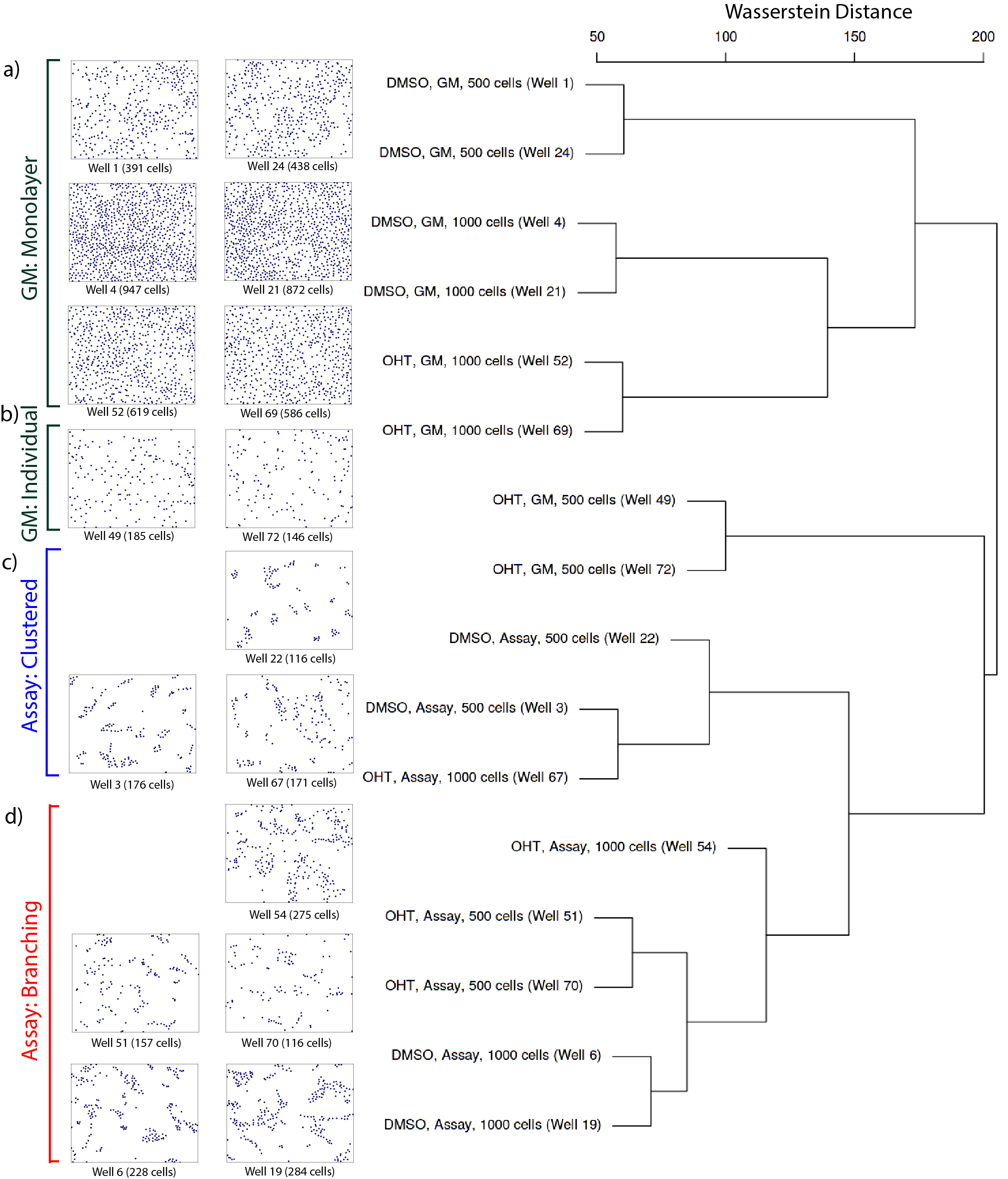}
\caption{\textbf{Classification of experimental conditions based on pairwise Wasserstein distance groups similar experimental conditions (e.g. cell density, biochemical treatment)}. ``DMSO'' treatment corresponds to an epithelial phenotype, ``OHT'' treatment corresponds to  an induced EMT phenotype, ``GM'' correspnds to growth media with 20 ng/mL EGF, ``Assay'' corresponds to assay media with 0.075 ng/mL EGF. Cells were seeded at initial densities of 500 or 1000 cells per well.}
\label{fig:Figure5}\vspace{-0.2in}
\end{figure*}

Hierarchical clustering based on pairwise Wasserstein distances between cell nuclei positions first partitioned the experimental groups that were cultured in growth media (20 ng/mL EGF) or assay media (0.075 ng/mL EGF), respectively (\textbf{Fig.~\ref{fig:Figure5}}). Cells cultured in growth media conditions were typically highly motile and proliferative, and could be further classified as dense monolayers  (\textbf{Fig.~\ref{fig:Figure5}a}) or dispersed individuals (\textbf{Fig.~\ref{fig:Figure5}b}). At the lowest level, replicate experiments with comparable biochemical treatments and initial cell densities were also grouped together, indicating their high similarity. Interestingly, the OHT-treated conditions with growth media and lower initial cell density (500 cells/well) were classified separately from the other growth media conditions, and appeared individual (\textbf{Fig.~\ref{fig:Figure5}c}). This is consistent with the effect of this biochemical treatment, since OHT-treatment to induce Snail and EMT resulted in enhanced motility, slower proliferation, and downregulated cell-cell junctions, particularly at lower starting cell densities.

In comparison, cells cultured with assay media (0.075 ng/mL EGF) exhibited lower cell densities after 60 h, and were further grouped into clustered or branching phases. For instance, DMSO-treated cells at lower initial cell density (500 cells/well) typically organized into morphologically compact clusters that were spatially separated  (\textbf{Fig.~\ref{fig:Figure5}c}). In comparison, OHT-treated cells were grouped together and displayed branching, dendritic architectures at both initial cell densities (500, 1000 cells/well), consistent with our previous results (\textbf{Fig.~\ref{fig:Figure5}d}).\cite{Leggett:2019} Finally, DMSO-treated cells at higher initial cell densities (1000 cells/well) also formed branching, dendritic architectures (\textbf{Fig.~\ref{fig:Figure5}d}). It should be noted that our analysis is based on the cell nuclear positions only, whereas the cell morphology in the experiments was highly elongated. Thus, cells in experiments could be connected together into spanning networks over longer distances than a typical epithelial cell length.

Finally, we compared experimental conditions cultured with assay media (0.075 ng/mL EGF) relative to treatment with gefitinib (500 nM), which inhibits downstream signaling of the EGFR pathway.\cite{Barr:2008cq} Our previous experiments showed that gefitinib treatment results in qualitatively similar branching configurations, albeit with slightly faster proliferation relative to assay media. Hierarchical clustering grouped experimental conditions by assay media or by gefitinib treatment, respectively (\textbf{Fig.~\ref{fig:SI_Fig8}}). In assay media, cells typically organized as branching morphologies with elongated arms of single-file cells (\textbf{Fig.~\ref{fig:SI_Fig8}a}). One OHT and gefitnib treated condition (500 cells/well) was grouped with the other assay media conditions, but appeared more consistent with these sparse branching network morphologies by visual inspection. In comparison, gefinib treatment also resulted in branching morphologies with arms that were many cells wide (\textbf{Fig.~\ref{fig:SI_Fig8}b}). One OHT and assay media condition (1,000 cells/well) was grouped in with the other gefitinib treated conditions, and also appeared consistent with these wider branching morphologies by visual inspection.

Based on the qualitatively similar appearance of our experiments and simulations, we then used Wasserstein distance to classify representative experimental conditions with the most similar simulation conditions. Experimental conditions cultured using growth media (20 ng/mL EGF) were classified within the branching phase with high self-propulsion force (Well 1 and Well 24), consistent with their increased motility and proliferation (\textbf{Fig.~\ref{fig:SI_Exp_Phase_Diagram}a,b}). We note that these two conditions were plated at a lower initial cell density, and only formed a subconfluent monolayer at the completion of the experiment. Next, the gefitnib treated conditions were located near the phase boundary (Well 23 and Well 53), since their inhibition of EGFR signaling resulted in somewhat reduced motility and proliferation. Notably, gefitinib and OHT treatment (Well 53) resulted in more branched structures due to EMT induction (\textbf{Fig.~\ref{fig:SI_Exp_Phase_Diagram}a,b}). In comparison, gefitinib treatment alone (Well 23) resulted in more compact clusters, consistent with stronger cell-cell adhesions in the epithelial state (\textbf{Fig.~\ref{fig:SI_Exp_Phase_Diagram}a,c}). The assay media condition (0.075 ng/mL EGF) was also classified as clustered (Well 22) (\textbf{Fig.~\ref{fig:SI_Exp_Phase_Diagram}a,c}), with even lower propulsion, which can be explained by the minimal EGFR signaling in assay media. Overall, hierarchical clustering with TDA is able to distinguish clusters with varying morphology due to different biochemical treatments, which can be quantitatively mapped to topologically similar phases in the self-propelled particle simulations.

\subsection{Topological Differences in Particle Configuration Depends on the Presence and Persistence of Loops}

\begin{figure}
\includegraphics[width=8.3cm]{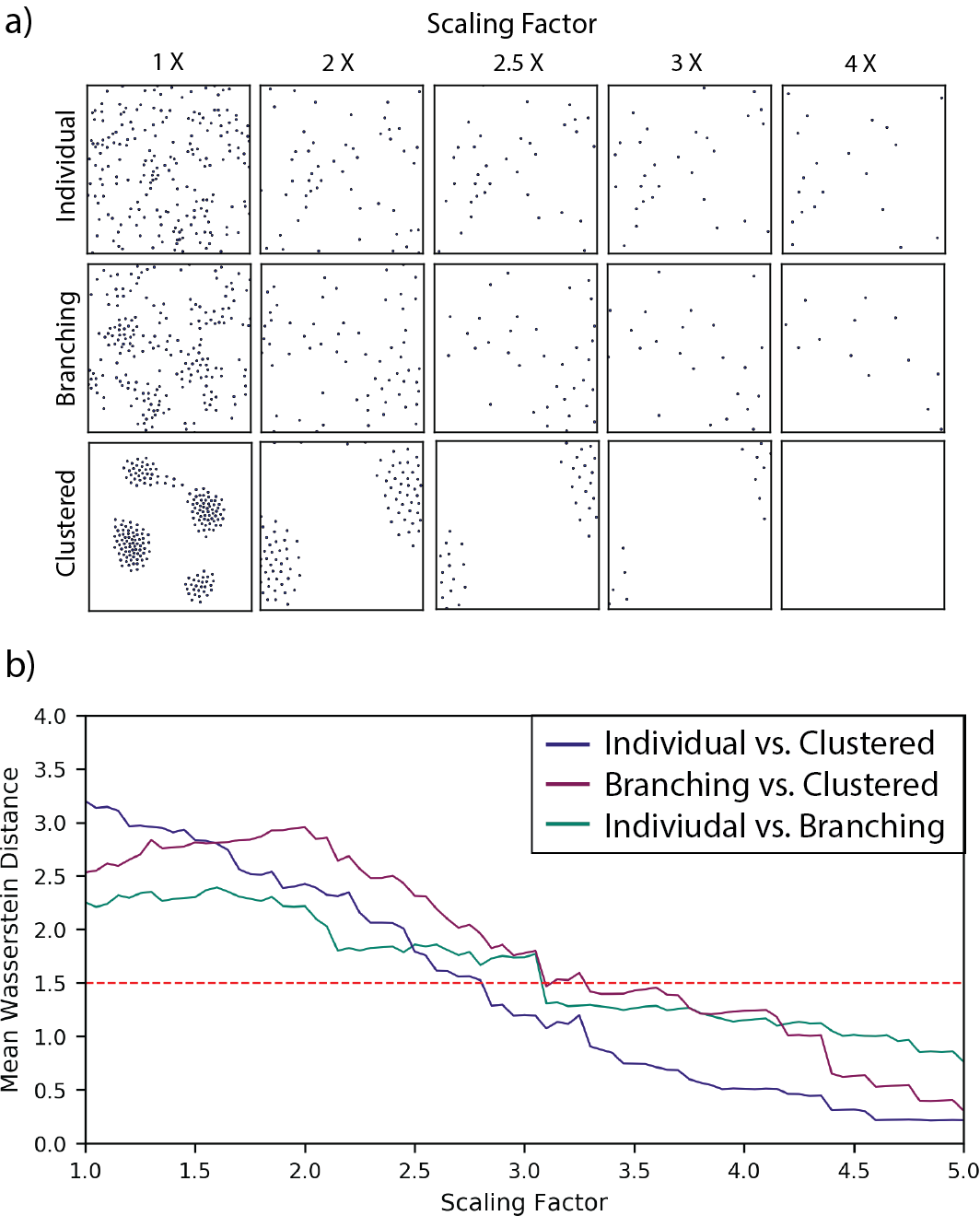}
\caption{\textbf{Wasserstein distance with varying spatial scaling for representative simulations at constant population size.} a) Representative snapshots of particle configuration within progressively smaller regions of interest (1X-4X). b) Mean Wasserstein distances between $10$ random initializations of individual ($\alpha = 0.07, P = 0.021$), branching ($\alpha = 0.09, P = 0.011$), or clustered phases ($\alpha = 0.23, P = 0.009$) at varying spatial scaling.}
\label{fig:spatial_scaling}
\end{figure}

For a given particle configuration, the presence of topological loops sets a length scale which is significantly larger than the particles themselves. This suggests that the topological differences between particle configurations are most apparent when sampling over some length scale larger than the size of the topological loops. We explored this scale-dependence by considering 10 representative individual, branched, and clustered configurations, and computing the Wasserstein distance between them for subsets of particles within progressively smaller regions of interest (ROI) (\textbf{Fig.~\ref{fig:spatial_scaling}a}). In the full field of view (1X scaling), the Wasserstein distance was greatest when comparing individual to clustered configuration ($\approx 3$), which were the most topologically different in terms of loop number and size (\textbf{Fig.~\ref{fig:spatial_scaling}b}).  In comparison, Wasserstein distance was smaller when comparing branching to individual to clustered configurations ($\approx 2$), which were more topologically similar. In practice, we defined a cutoff Wasserstein distance of 1.5, below which particle configurations were considered similar and above which particle configurations were dissimilar. For progressively smaller fields of view (2-3X), some of the particles forming the closed loop were ``cropped out,'' decreasing the persistence or even destroying the large topological loops. As a consequence, the topological differences between clustered phases with individual or branching phases became less pronounced. Indeed, when the field of view was appreciably smaller than the loop size (3X onward), the clustered and individual phases became the least topologically different relative to the others. These results indicate that TDA-based classification based on loops is most effective when sampling over length scales larger than the loops.

\begin{figure}
\includegraphics[width=8.3cm]{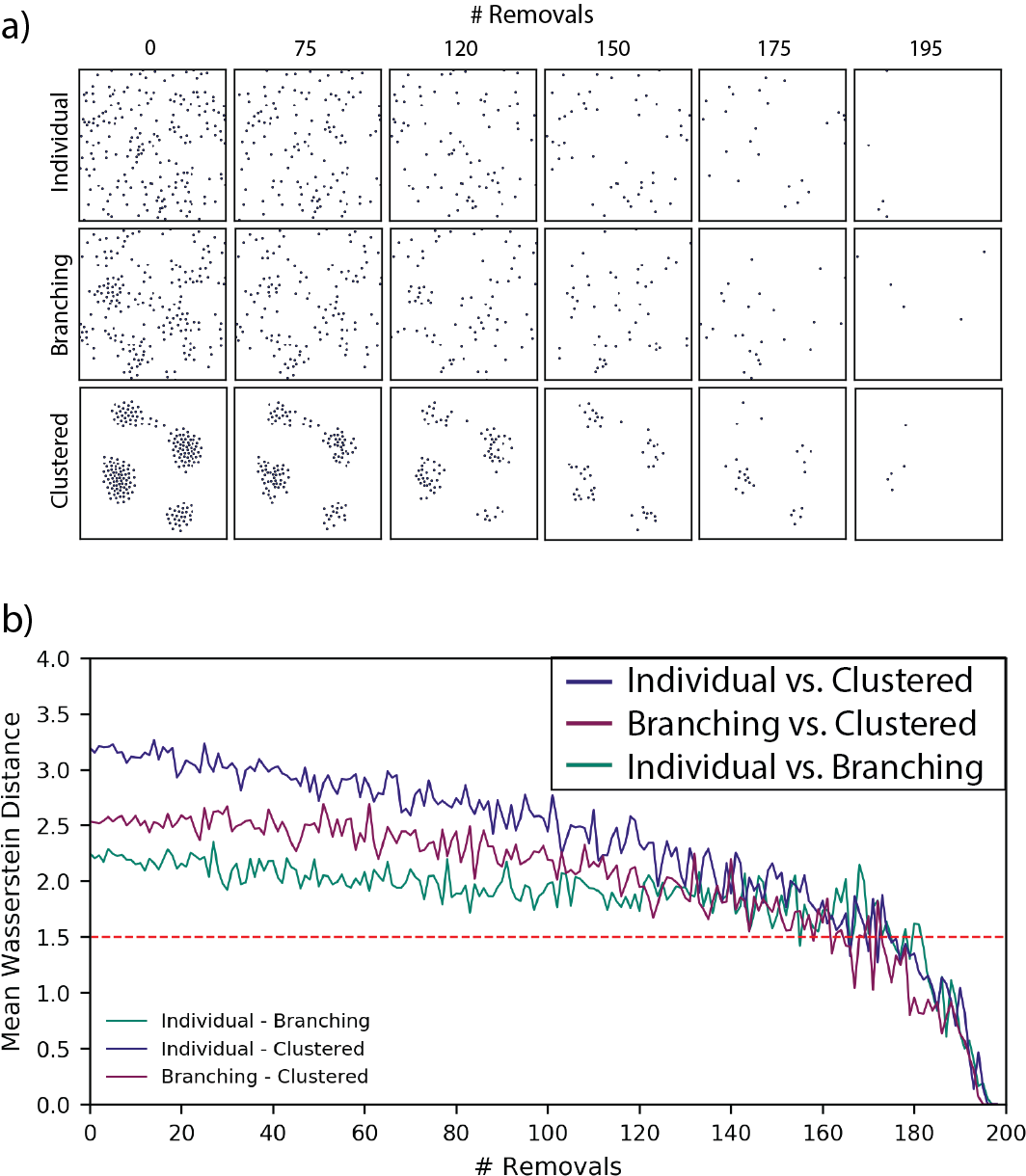}
\caption{\textbf{Wasserstein distance with random particle removal for representative simulations at constant population size.} a) Representative snapshots of particle configuration with increasing random particle removals. b) Mean Wasserstein distances between $10$ random initializations of individual ($\alpha = 0.07, P = 0.021$), branching ($\alpha = 0.09, P = 0.011$), or clustered ($\alpha = 0.23, P = 0.009$) phases with increasing particle removal.}
\label{fig:removal}
\end{figure}

Alterations in the spatial connectivity of adjacent particles would also be expected to affect the persistence of large topological loops. For instance, the random removal of particles within a loop could increase the minimum distance needed to connect all the points, also reducing the persistence of the loop. To systematically investigate how this sparser sampling would affect topological loops, we again considered 10 representative individual, branched, and clustered phases, randomly removed particles, and recomputed the Wasserstein distance (\textbf{Fig.~\ref{fig:removal}a}). Remarkably, we find that the Wasserstein distance decreases relatively slowly out to 100 particle removals (50\%) (\textbf{Fig.~\ref{fig:removal}b}). Indeed, the Wasserstein distances do not fall below the cutoff of 1.5 until 160 particle removals (80\%). This result is plausible based on the representative snapshots of particle configuration, where the underlying patterns are still apparent up when 175 particles have been removed (88\%) (\textbf{Fig.~\ref{fig:removal}a}). We speculate that the disappearance of any loop roughly occurs when the spacing between particles surrounding the loop becomes comparable to the loop size. In actuality, this allows for extremely sparse representations of the loop with only 40 remaining particles (20\%). This result shows that incomplete or even sparse sampling of the particle configuration would most likely be adequate to classify topological differences between them.

As a proof of concept, we further applied this approach to classify different particle configurations over time, for fixed population size (\textbf{Fig.~\ref{fig:SI_Fig3}}), as well as proliferating particles (\textbf{Fig.~\ref{fig:SI_Fig6}}). For fixed population size, the mean Wasserstein distance between different phases typically increased over time as particles aggregated into branching or compact clusters, and exceeded the threshold of $1.5$ at $t < 500$ (within $25,000$ timesteps, \textbf{Fig.~\ref{fig:SI_timelapse_wass}a}). All three mean Wasserstein distances approached a plateau value by $t = 1000$ ($50,000$ timesteps), indicating that the topological configuration was persistent over the remainder of the simulation (\textbf{Fig.~\ref{fig:SI_timelapse_wass}a}). For comparison, the nearest neighbor counts for individual and branching simulations also reached a plateau value by $t = 500$ (\textbf{Fig.~\ref{fig:SI_Fig12}a}), in agreement with the increase in mean Wasserstein distance above $1.5$ for their respective comparisons. Qualitatively similar trends were observed for proliferating particles, where the mean Wasserstein distance increased above $1.5$ by $t = 500$ (\textbf{Fig.~\ref{fig:SI_timelapse_wass}b}), in agreement with the timescale where the nearest neighbour counts diverge (\textbf{Fig.~\ref{fig:SI_Fig12}b}). Therefore, the TDA-based unsupervised classifier thus correctly identifies the time at which simulations become topologically dissimilar (e.g. undergoing a phase transition from individuals to clusters).

\section{Discussion and Conclusion}

We demonstrate unsupervised classification of collective and individual phases in epithelial cells based on the persistence of topological loops (i.e. dimension 1 homology). This approach incorporates spatial information at short length scales (by identifying the proximity of a particle to its neighbors), as well as at longer length scales (by identifying how particles can be linked together into a closed loop around an empty region). The loop size sets a characteristic length scale for the system, so that sampling particle configurations within a region of interest smaller than this length scale reduces classification accuracy. Nevertheless, these topological loops are highly persistent, since sampling some subset of the particles appears sufficient to define the loop (that is, an underlying one-dimensional manifold). Iterative removal of particles at random from the configuration and reclassification shows that loops remain even when nearly all the particles are removed ($80\%$). We estimate that loops can be recovered until the spacing between points becomes comparable to the loop diameter. In comparison, classification using connected components only accounts for local information and is analogous to local order parameters. We show that classification using connected components (dimension 0 homology) yields qualitatively similar trends but is skewed towards higher density or numbers relative to connected loops (dimension 1 homology) (\textbf{Fig.~\ref{fig:SI_Fig_H0_phase}, ~\ref{fig:SI_Fig_noprolif_H0}, ~\ref{fig:SI_Fig_prolif_H0}}). In the work of others, the use of connected components appears to require arbitrary normalization to account for differences in population size,\cite{atienza2019persistent} or otherwise is weighted by local density and population size.\cite{skinner2020topological}

This work primarily focused on particle configuration at fixed snapshots, and did not explicitly consider dynamics. As a proof-of-concept, we showed that different phases can be classified for different particle configurations at varying times. Nevertheless, temporally varying topological barcodes have been previously demonstrated elsewhere (at constant particle number),\cite{Topaz:2015gd,Ulmer:2019hx,Bhaskar:2019hj} and could be implemented to provide additional insights into particle dynamics. Indeed, TDA could enable efficient sampling of time-series data to identify events of interest across varying simulation parameters. In the future, we envision that TDA could be generalized across different types of propulsion mechanisms and interparticle interactions to infer unifying principles for self-organization.\cite{Cichos:2020ce}

The total computational time for unsupervised classification using TDA was on the order of minutes using a multicore CPU, and was not limiting for these simulations (\textbf{Note S3}). More generally, the computational cost does not exhibit trivial scaling with particle number, since it varies with the local particle configuration with respect to empty space. As an illustrative example, it will be more computationally expensive to construct the simplicial complex for 100 densely packed points roughly organized as a single loop around an empty region, relative to 10 well separated loops of 10 points each, arranged in a single file. Moreover, for particles arranged in three-dimensional space, we expect that computation of higher-order structures around empty voids (i.e. dimension 2 homology) will require additional bookkeeping with increased memory requirements. If the construction of a simplicial complex becomes prohibitively expensive for an extremely large number of particles, we anticipate that some sparse sampling will likely be adequate to recover topological loops and voids (i.e. lazy witness complex). Further, the computation of Wasserstein distance between persistence diagrams with persistent loops (dimension 1 homology) tends to be significantly faster than for connected components (dimension 0 homology). This is unsurprising since there tend to be fewer persistent loops than persistent connected components, which reduces the complexity for finding the optimal matching to compute Wasserstein distance. The pairwise computation of Wasserstein distance between pairs of simulations was implemented in parallel, and we also anticipate this could be completed in a reasonable time on a cluster.

The minimal model of epithelial cells as self-propelled particles neglects interesting biological mechanisms that also drive collective migration. For instance, this model does not consider cell shape changes,\cite{Leggett:2016fz} which can affect cell-cell interactions as well as motility. Moreover, this model does not address the sensing or release of soluble biochemical signals, which can also function to recruit or repel cells through directed migration.\cite{Camley_2018} One crucial question is whether a population of cells can truly be treated as homogeneous, due to genetic and non-genetic heterogeneity that is manifested at the single cell level.\cite{Altschuler:2010ky} Indeed, mixtures of two different cell types can exhibit fascinating self-sorting behaviors, which would not be observed with either cell type alone.\cite{Mehes:2012fy,GamboaCastro:2016da} Moreover, cells may alter their migration phenotype over time, such as a epithelial-mesenchymal transition from clustered epithelial cells to individual mesenchymal cells.\cite{Wong:2014jma} There is extensive interest in the emergence of ``leader cells'' that exhibit a partial EMT, allowing collective guidance of mechanically connected followers.\cite{Reffay:2014ku,Vishwakarma:2018cl,Leggett:2019} The application of TDA to elucidate biological heterogeneity in an experimental and computational context also represents a fruitful direction for further work.

In conclusion, we demonstrate topology-based machine learning to classify spatial patterns of epithelial cells in a robust and unbiased fashion. We validate this unsupervised classification using simulated data of self-propelled particles with varying adhesion, both at fixed population size and with contact-inhibited proliferation, exhibiting good agreement with local order parameters. We further show that experimental measurements of mammary epithelial cells group together by replicate conditions and biochemical treatments, and can be mapped back to simulations with physically plausible parameters. Finally, we explore the persistence of topological structure at varying length scales, sparse sampling, and over time, revealing why classification is highly robust. Overall, this approach reveals that transitions between individual and collective phases can be identified based on differences in particle configuration around empty regions, which should be generic and widely applicable to a variety of active and biological systems at multiple scales.

\section*{Conflicts of interest}
There are no conflicts to declare.

\section*{Acknowledgements}
We thank S.E. Leggett and Z.J. Neronha for acquiring and processing the experimental data used in this manuscript, which was previously published elsewhere.\cite{Leggett:2019} This work was supported by the National Cancer Institute's Innovative Molecular Analysis Technologies (IMAT) Program (R21CA212932) and a Brown University Data Science Initative Seed Grant. This research was conducted using computational resources and services at the Center for Computation and Visualization, Brown University.

\balance
\bibliography{references}
\bibliographystyle{references}

\newcommand{\SupportingPrefix}{S}
\renewcommand{\thefigure}{\SupportingPrefix\arabic{figure}}
\renewcommand{\thetable}{\SupportingPrefix\arabic{table}}
\setcounter{figure}{0}
\setcounter{table}{0}
\setcounter{page}{1}
\onecolumn
\section{Electronic Supplemental Information} \label{Supp}

\section*{Note S1: Computational of Persistence Diagrams and Wasserstein Distance}
\begin{figure*}[h!]
\centering
\includegraphics[width=7.64cm]{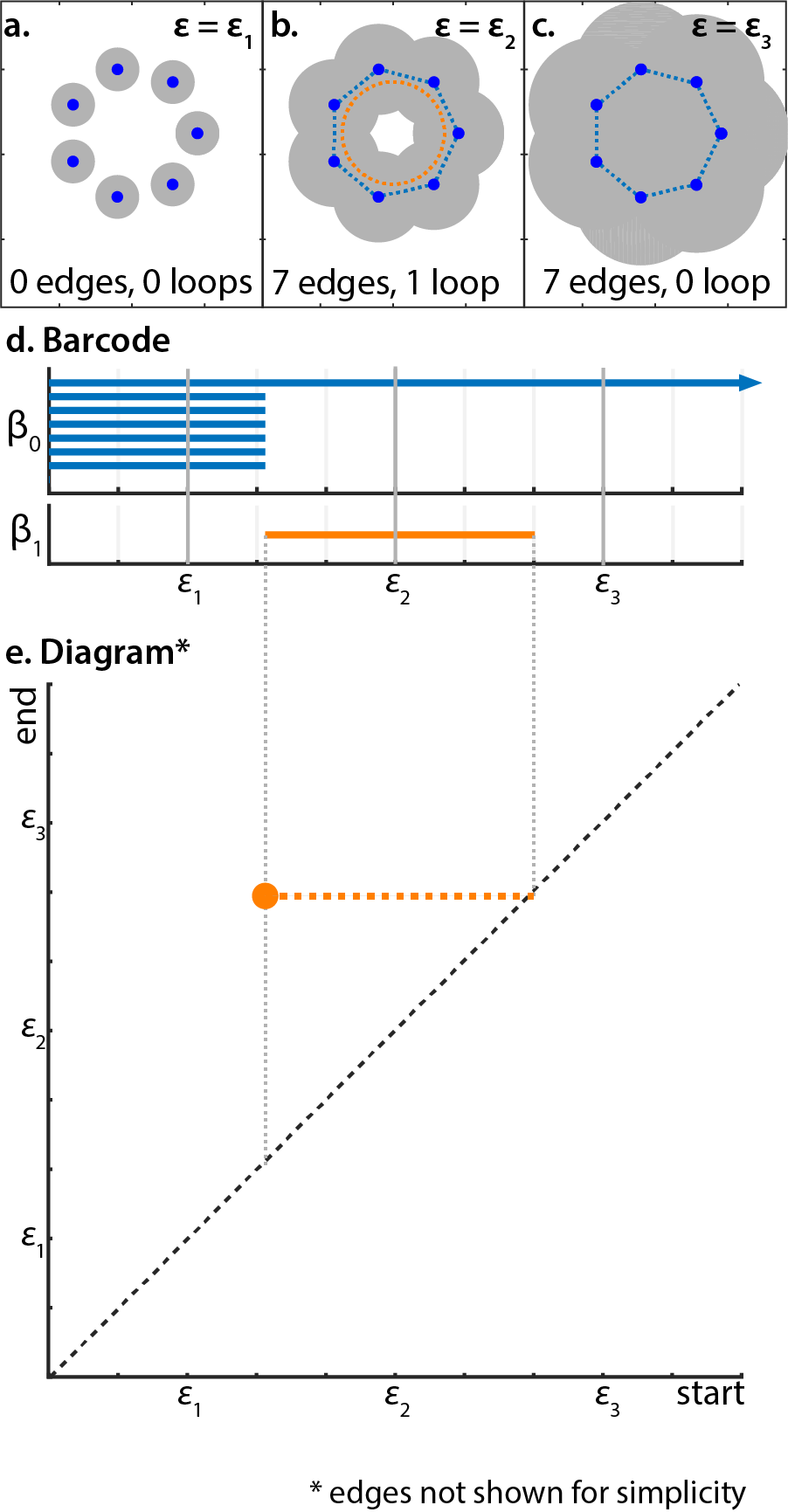}
\caption{\textbf{Computation of persistence homology.} (a,b,c) Visualization of connectivity between points at varying values of spatial parameter $\epsilon$. Edges are colored in blue (dimension 0 homology), and loops are colored in orange (dimension 1 homology) (d) Corresponding topological barcode. (e) Persistence diagram showing loops only for simplicity.}
\label{fig:barcodediagram}\vspace{-0.2in}
\end{figure*}

Topological barcodes visualize the robustness (via persistent homology) of topological features such as edges and loops across varying length scales (i.e. filtration values) $\epsilon$.\cite{Ghrist:2008bi} For example, consider a set of 7 points at varying filtration, illustrated by gray disks with radius $\epsilon_1$ centered at each point (\textbf{Fig.~\ref{fig:barcodediagram}a}). As the radius increases to $\epsilon_2$, certain gray disks overlap, indicating that the corresponding points should be connected by 7 edges (blue lines, dimension 0 homology) at this cutoff distance (\textbf{Fig.~\ref{fig:barcodediagram}b}). Moreover, these connected edges form 1 closed loop around an empty region (orange circle, dimension 1 homology). A further increase in radius to $\epsilon_3$ maintains connectivity of the 7 edges, but collapses the closed loop (\textbf{Fig.~\ref{fig:barcodediagram}c}).

\vspace{1cm}

The corresponding topological barcode uses horizontal bars to visualize the persistence of features that appear or disappear at varying $\epsilon$ intervals (\textbf{Figure~\ref{fig:barcodediagram}d}). Essentially, the barcode presents Betti intervals whose start corresponds to the $\epsilon$-value for the appearance of a topological feature (i.e. the lowest $\epsilon$-value at which the feature appears (``start''), and whose end corresponds to the $\epsilon$-value for the disappearance of the topological feature (i.e. the highest $\epsilon$-value at which the feature is present,). For example, at $\epsilon_1$, there are seven distinct blue bars corresponding to the number of discrete points, which are not connected at this length scale. At $\epsilon_2$, there is only one blue bar, since all points have linked together forming a connected component, and there is the appearance of one closed loop, indicated by an orange bar. At $\epsilon_3$, there remains only one blue bar and no orange bar, since the closed loop has collapsed.  The same information can also be organized in a persistence diagram (start $\epsilon$ values on x-axis and end $\epsilon$ values on y-axis), where the distance from the diagonal is indicative of the significance or importance of a topological feature (\textbf{Fig.~\ref{fig:barcodediagram}e}). In particular, note that the closed loop is appreciably offset from the diagonal, signifying that it is a relatively stable topological structure for this particle configuration. For simplicity, edges are not shown in the persistence diagram.\\

\begin{figure}[h!]
\centering
\includegraphics[width=8.3cm]{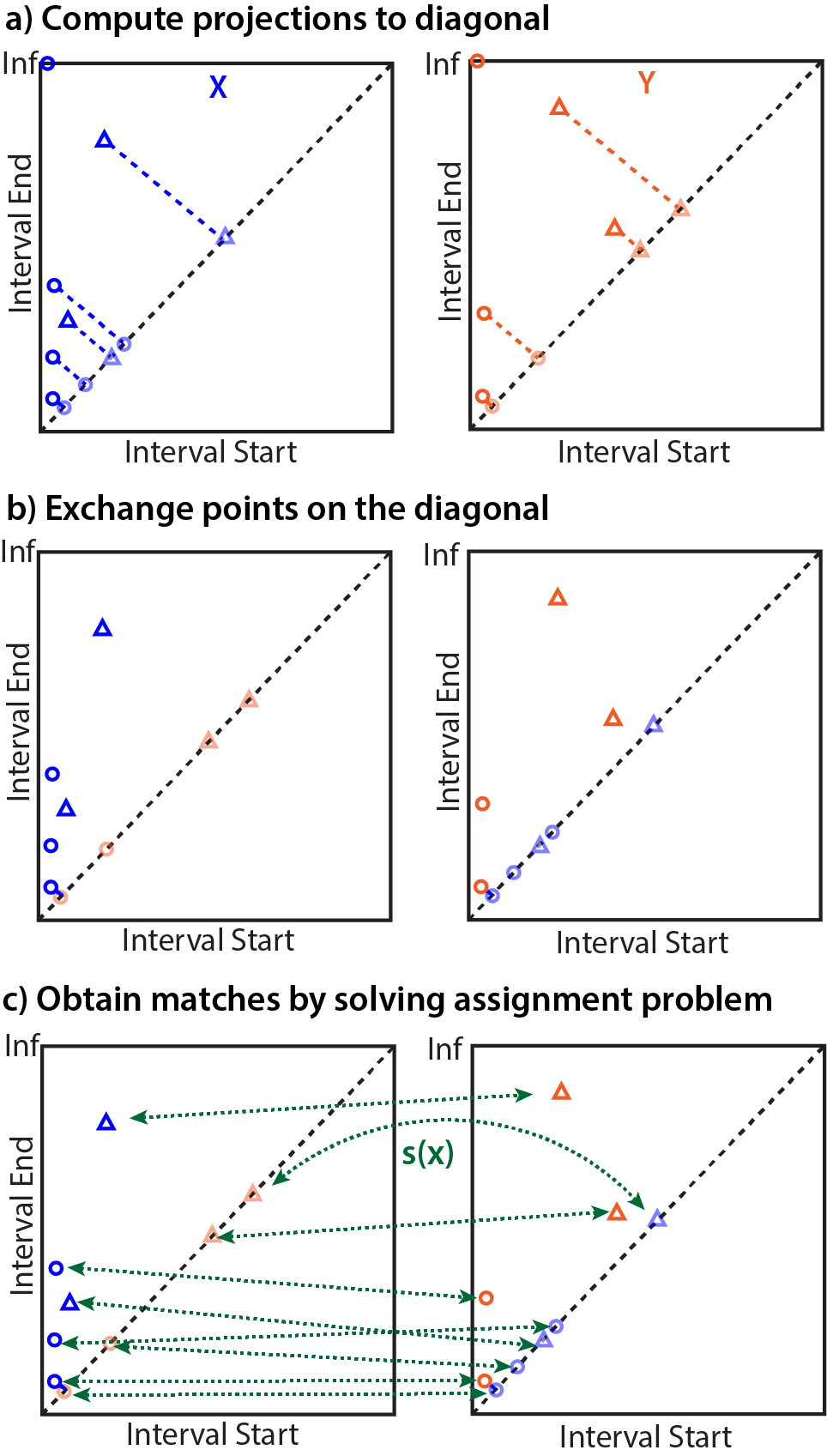}
\caption{\textbf{Computation of Wasserstein distance.} (a) Projections of off-diagonal points to the diagonal are computed. Circles represent edges or connected components and triangles represent loops. Point at $(0,\infty)$ representing $1$ connected component for high values of spatial parameter $\epsilon$ is not considered. (b) Projections on the diagonal are exchanged between persistence diagrams. (c) Points are matched to their closest neighbor in the other diagram. Note that points can also be matched to their diagonal projection. Circles can only be matched to other circles and triangles can only be matched to other triangles.}
\label{fig:SI_Fig2}\vspace{-0.2in}
\end{figure}

\begin{figure*}[h!]
\centering
\includegraphics[width=8cm]{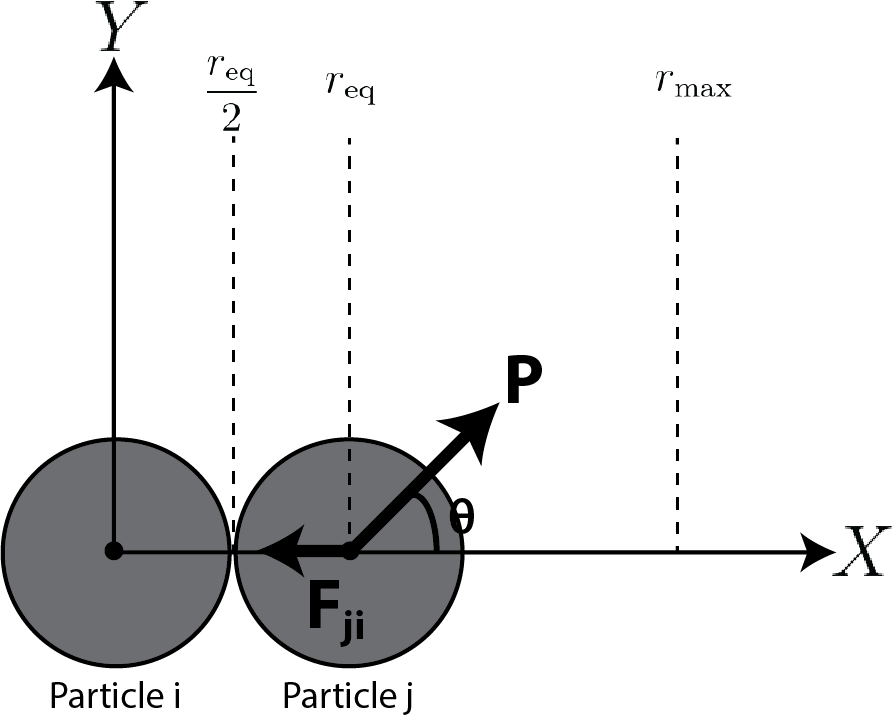}
\caption{\textbf{Schematic for particle-particle interaction in the agent-based model.} A particle $j$ is at equilibrium with particle $i$. A polarization force $P$ is applied to pull particle $j$ to distance $r_\text{max}$ against adhesion force $F_{ij}$.}
\label{fig:SI_FBD}\vspace{-0.2in}
\end{figure*}

\begin{figure*}[h!]
\centering
\includegraphics[width=18cm]{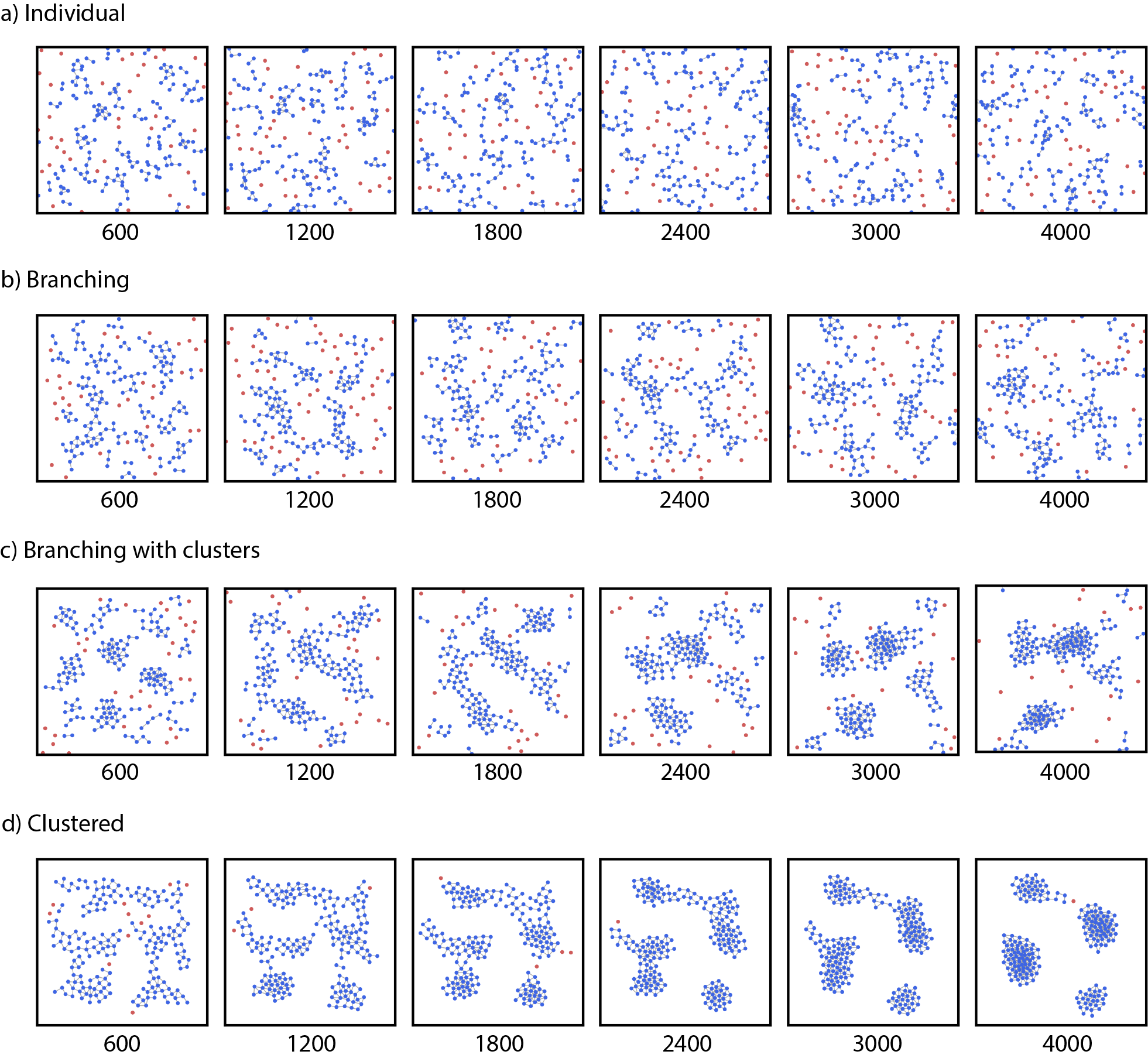}
\caption{\textbf{Non-proliferating particle model results in individual, branching, branching with clusters and clustered phenotypes with varying adhesion and self-propulsion force.} Representative snapshots at time intervals of 600 (every 30,000 timesteps, except for the last snapshot) of individual phase with $\alpha = 0.07, P = 0.021$ (a), branching phase with $\alpha = 0.09, P = 0.011$ (b), branching with clusters phase with $\alpha = 0.25, P = 0.017$ (c), and clustered phase with $\alpha = 0.23, P = 0.009$ (d), with all simulations starting with random initial positions. Particle with one or more neighbors are plotted in blue, with a ``bond'' drawn between any two cells within radial distance $1.0$. Individual cells are shown in red. More individual cells are observed when random polarization dominates over adhesion force. Furthermore, the presence of clusters at low adhesion is transitory and cells are highly motile.}
\label{fig:SI_Fig3}\vspace{-0.2in}
\end{figure*}

\begin{figure*}
\centering
\includegraphics[width=18cm]{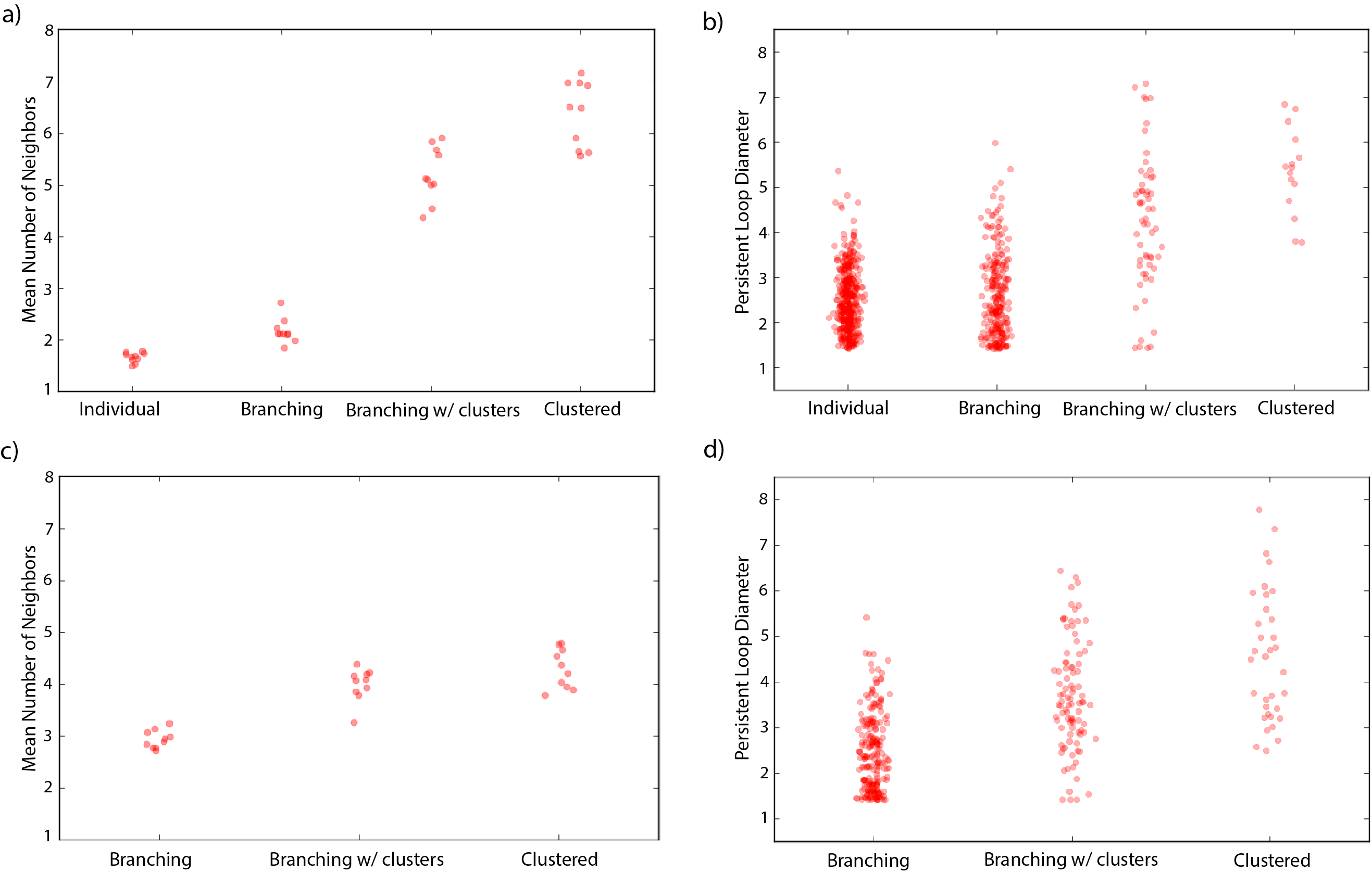}
\caption{\textbf{Comparison of local density (nearest neighbor count) and persistent loop diameter associated with distinct phases.} In the top row, data acquired from individual, branching, branching with clusters and clustered simulations performed $10$ times each with random initialization and parameters $(\alpha = 0.07, P = 0.021)$, $(\alpha = 0.09, P = 0.011)$, $(\alpha = 0.25, P = 0.017)$ and $(\alpha = 0.23, P = 0.009)$ respectively. In the bottom row, data acquired from branching, branching with clusters and clustered simulations performed $10$ times each with random initialization and parameters $(\alpha = 0.09, P = 0.009)$, $(\alpha = 0.19, P = 0.013)$ and $(\alpha = 0.23, P = 0.007)$ respectively. (a) Comparison of nearest neighbor count for individual, branching, and clustered phases at constant population size. (b) Comparison of characteristic loop diameter for individual, branching, and clustered phases at constant population size. (c) Comparison of nearest neighbor counts for branching and clustered phases in proliferating populations. (d) Comparison of characteristic loop diameter for individual, branching, and clustered phases in proliferating populations.}
\label{fig:SI_Fig4}\vspace{-0.2in}
\end{figure*}

\begin{figure*}
\centering
\includegraphics[width=14cm]{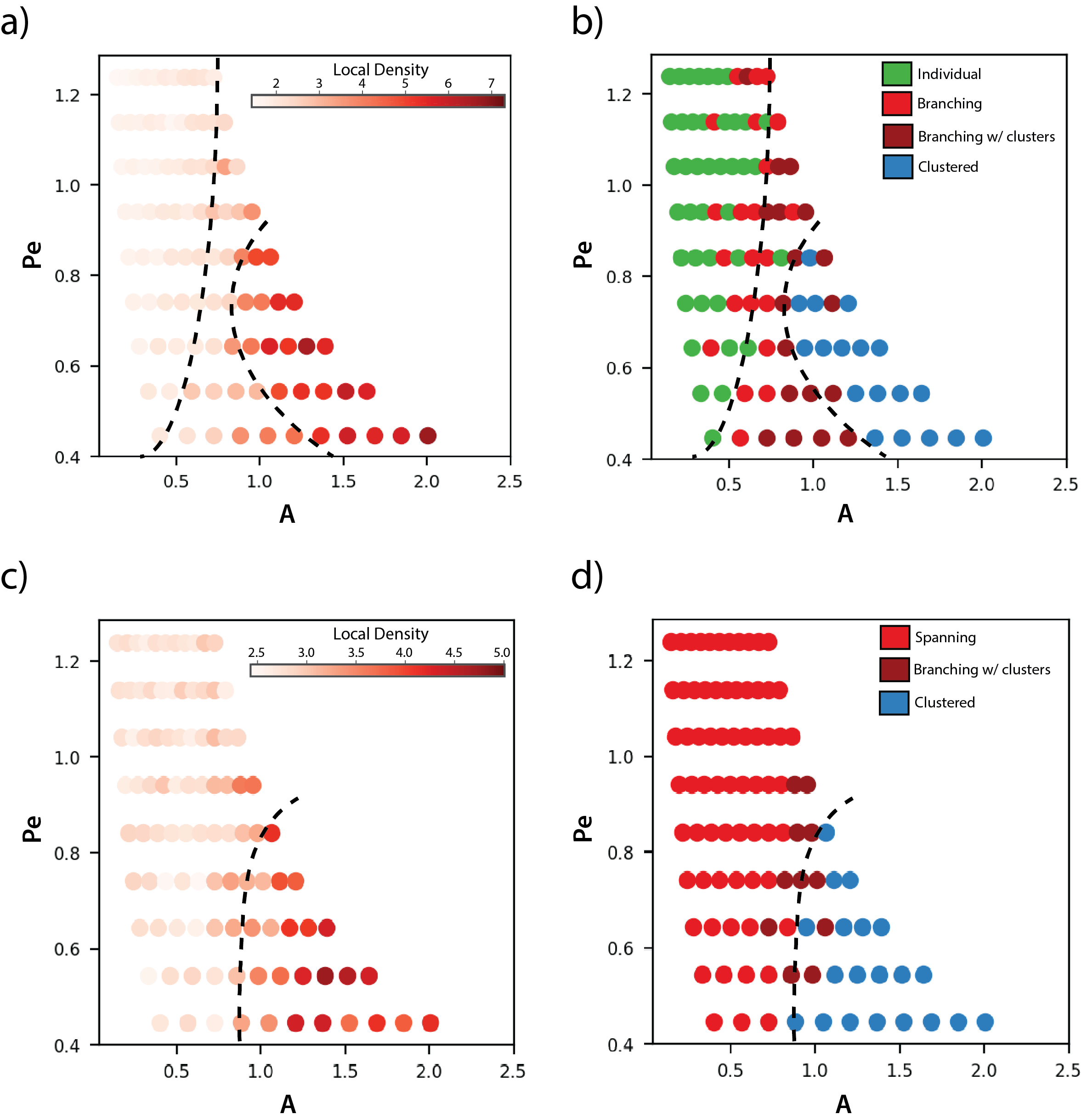}
\caption{\textbf{Phase transitions parameterized by dimensionless parameters for self-propulsion ($Pe$) and adhesion ($A$).} (a) Local density (nearest neighbor count) for a constant population size. (b) TDA classification of individual, branching, and clustered phases at constant population size. (c) Local density (nearest neighbor count) for a proliferating population. (d) TDA classification of branching and clustered phases for a proliferating population.}
\label{fig:SI_nondim_phase_diag}\vspace{-0.2in}
\end{figure*}

\begin{figure*}
\centering
\includegraphics[width=19cm]{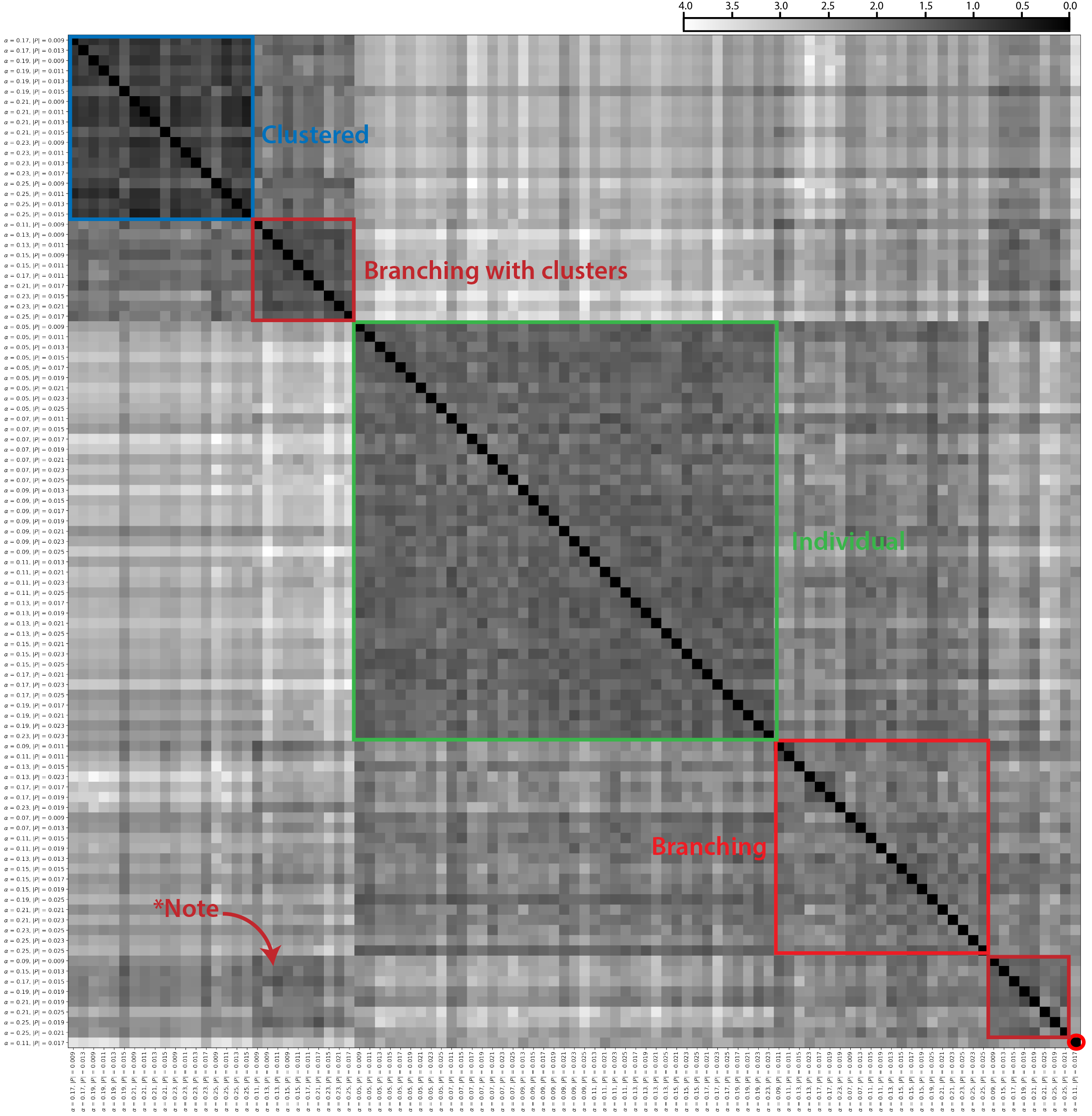}
\caption{\textbf{Pairwise Wasserstein distances between all simulations without proliferation, as well as varying adhesion and self-propulsion force.} Hierarchical clustering groups simulations into individual, branching, branching with clusters, and clustered categories corresponding to distinct phases along the diagonal. Note that branching with clusters phase exhibits similarity with both branching and clustered phases.}
\label{fig:SI_Fig5}\vspace{-0.2in}
\end{figure*}

\begin{figure*}
\centering
\includegraphics[width=18cm]{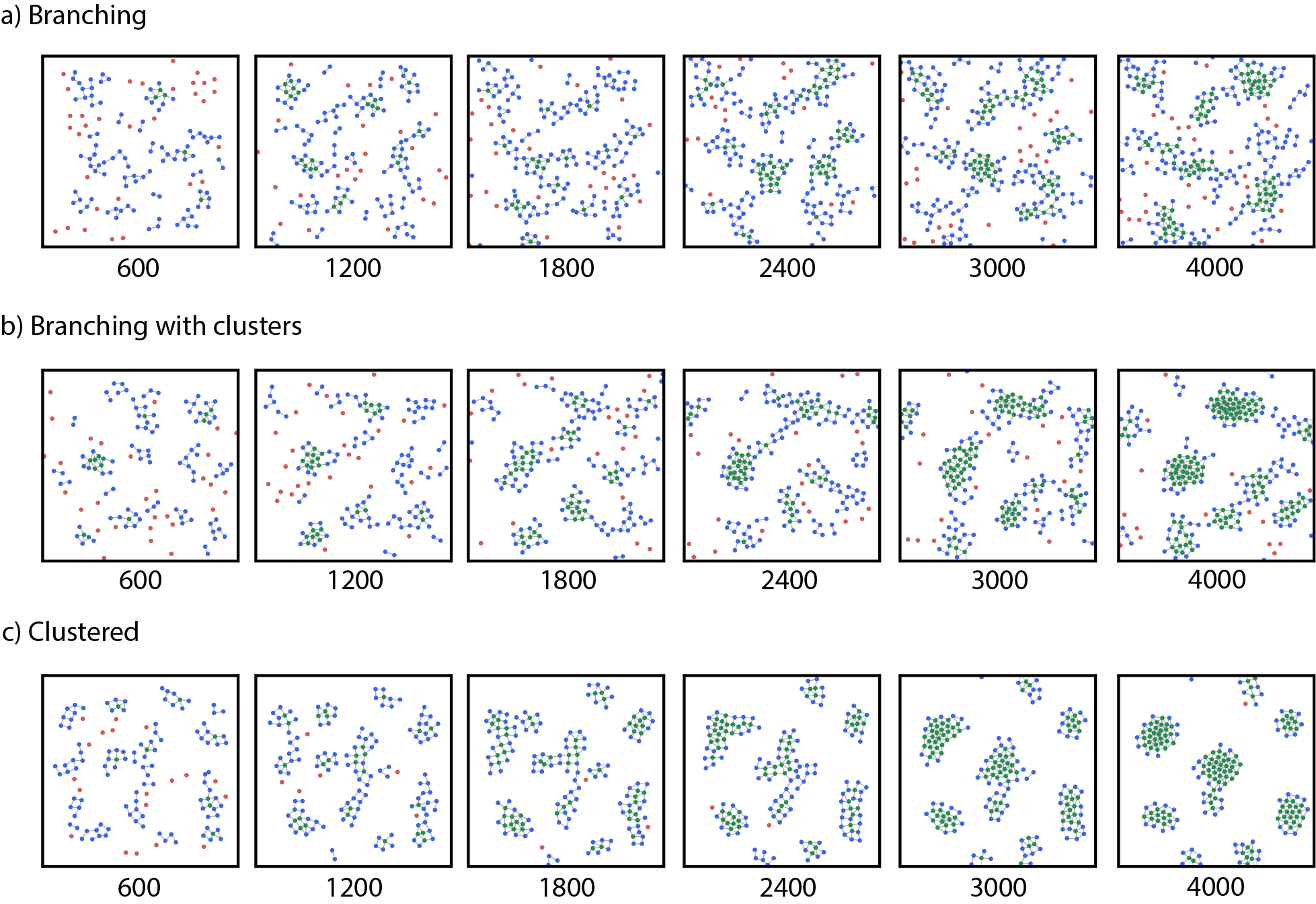}
\caption{\textbf{Self-propelled, proliferating particle model results in branching, branching with clusters and clustered phenotypes with varying adhesion and self-propulsion force.} Representative snapshots at time intervals of 600 (every 30,000 timesteps, except for the last snapshot) of branching phase with $\alpha = 0.09, P = 0.009$ (a), branching with clusters phase with $\alpha = 0.19, P = 0.013$ (b), and clustered phase with $\alpha = 0.23, P = 0.007$ (c), with all simulations starting with random initial positions. Particle with one or more neighbors are plotted in blue, with a ``bond'' drawn between any two cells within radial distance $1.0$. Particles with four or more neighbors that cannot proliferate due to contact inhibition of proliferation are shown in green.}
\label{fig:SI_Fig6}\vspace{-0.2in}
\end{figure*}

\begin{figure*}
\centering
\includegraphics[width=19cm]{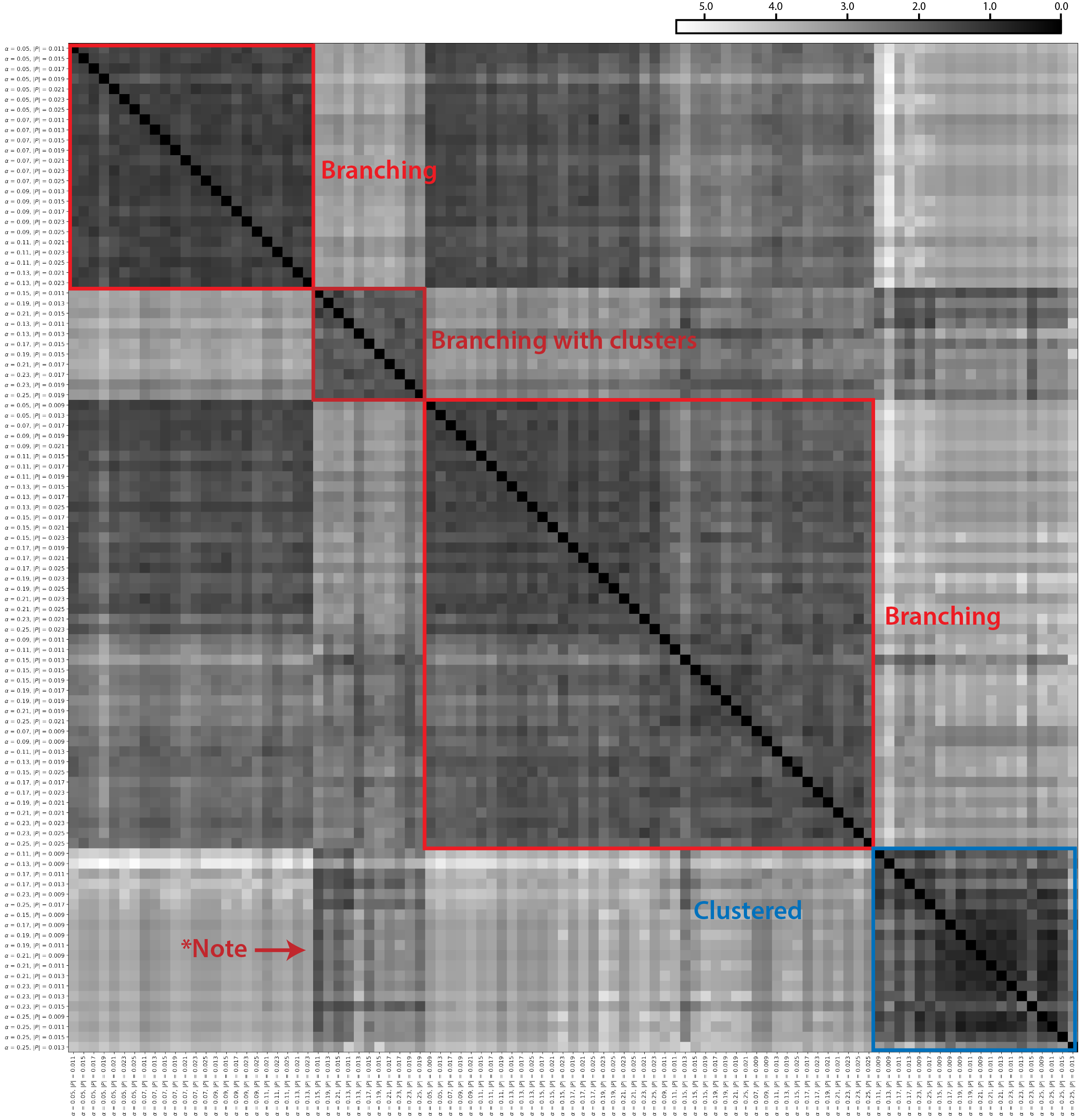}
\caption{\textbf{Pairwise Wasserstein distances between all simulations with proliferation, as well as varying adhesion and self-propulsion force.} Hierarchical clustering groups branching, branching with clusters, and clustered phases along the diagonal. Note that branching with clusters phase exhibits similarity with both branching and clustered phases.}
\label{fig:SI_Fig7}\vspace{-0.2in}
\end{figure*}

\begin{figure*}
\centering
\includegraphics[width=17.1cm]{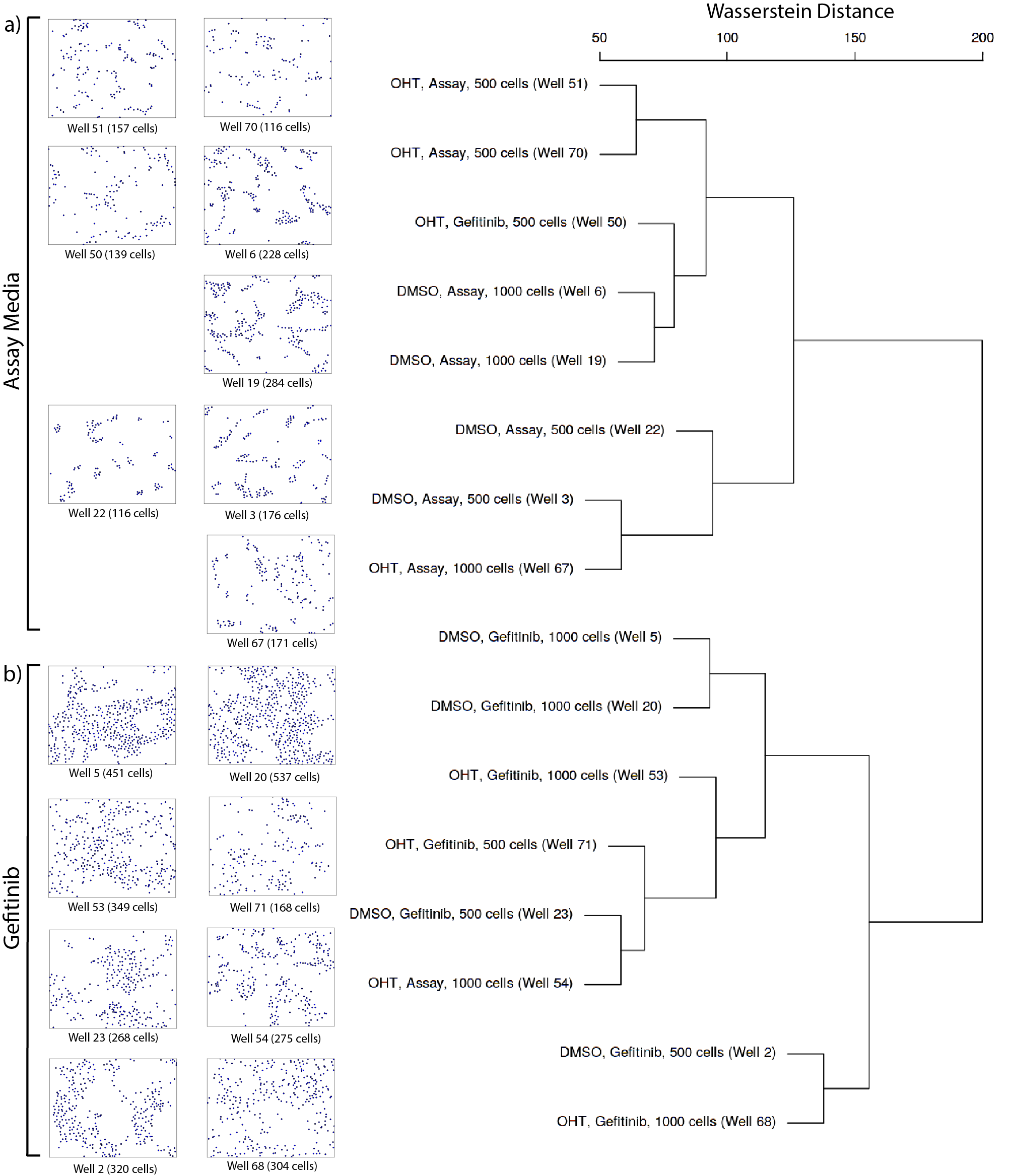}
\caption{\textbf{Hierarchical clustering of pairwise Wasserstein distances between persistence diagrams of experimentally measured cell nuclei positions identifies distinct clustered and branching phases.} Dendrogram obtained by running complete-linkage hierarchical clustering algorithm using Wasserstein distance groups experimental conditions based on assay media (a) and gefitinib treatment (b), then biochemical treatment with DMSO or OHT, as well as initial cell density.}
\label{fig:SI_Fig8}\vspace{-0.2in}
\end{figure*}

\begin{figure*}
\centering
\includegraphics[width=9cm]{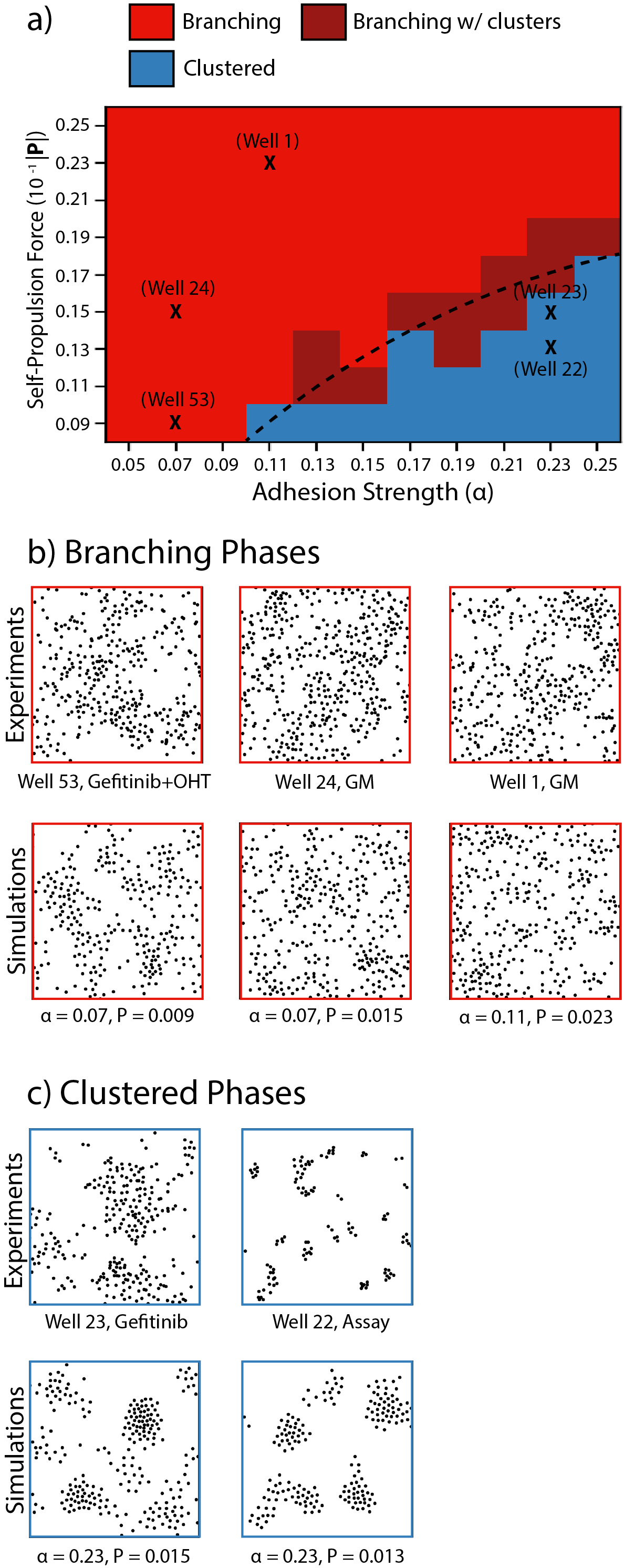}
\caption{\textbf{Experiments compared to simulation results using Wasserstein distance.} Pairwise Wasserstein distances are calculated between cell nuclei positions obtained from experiments (scaled to fit in $[-10,10]\times[-10,10]$ domain) and particle positions obtained from simulations. Experiments are matched to simulations by minimum distance. Experiments with Wasserstein distances exceeding $1.5$ are not shown.}
\label{fig:SI_Exp_Phase_Diagram}\vspace{-0.2in}
\end{figure*}

\begin{figure*}
\centering
\includegraphics[width=17cm]{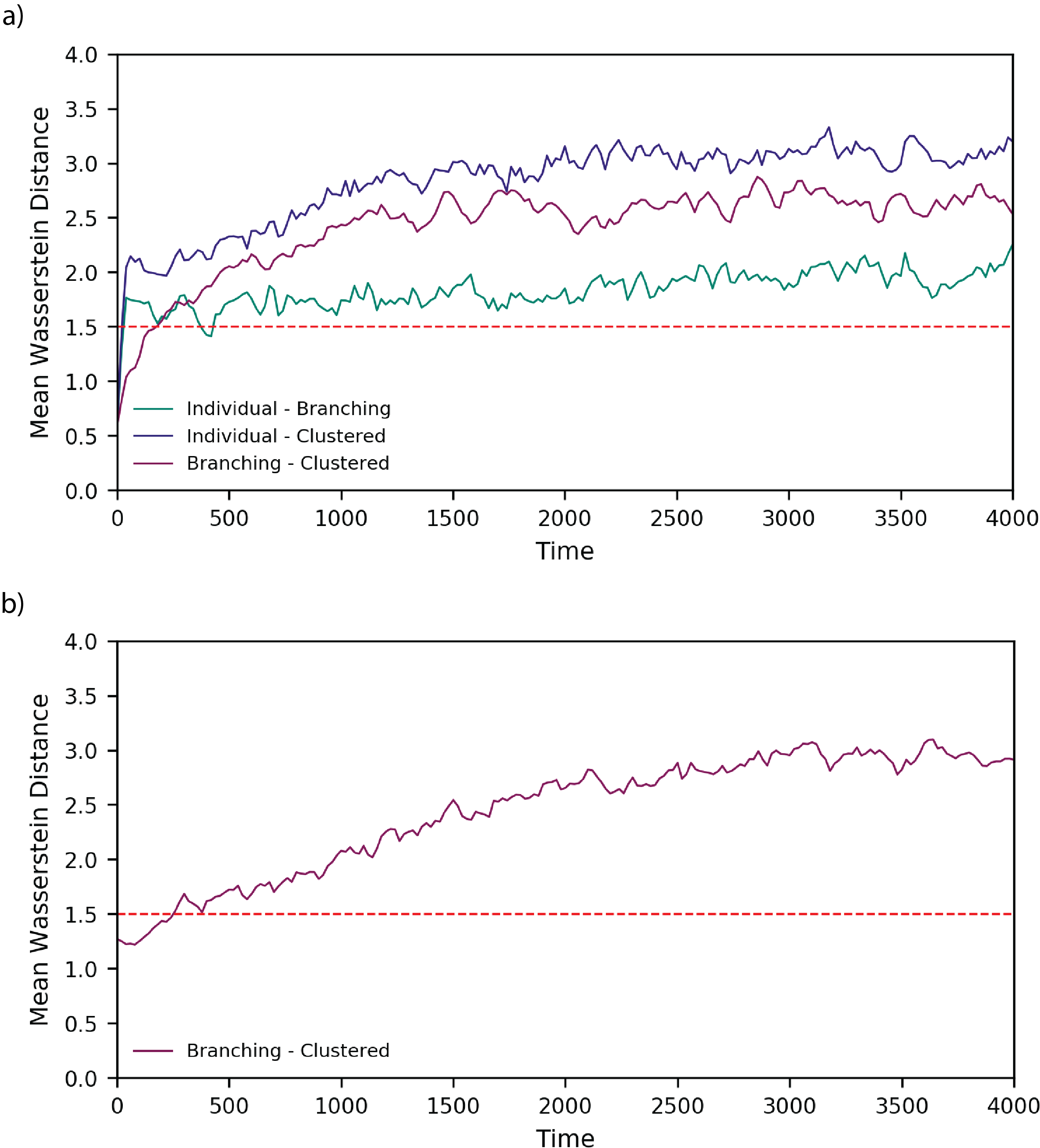}
\caption{\textbf{Pairwise Wasserstein distances over time.} (a) Pairwise Wasserstein distances comparing individual, branching, and clustered phases over time at fixed population size. (b) Comparison of branching, and clustered phases in a proliferating population. Dashed red line indicates threshold value above which the pair of simulations can be distinguished. Mean values computed using $10$ random initializations for each simulation.}
\label{fig:SI_timelapse_wass}\vspace{-0.2in}
\end{figure*}

\begin{figure*}
\centering
\includegraphics[width=17cm]{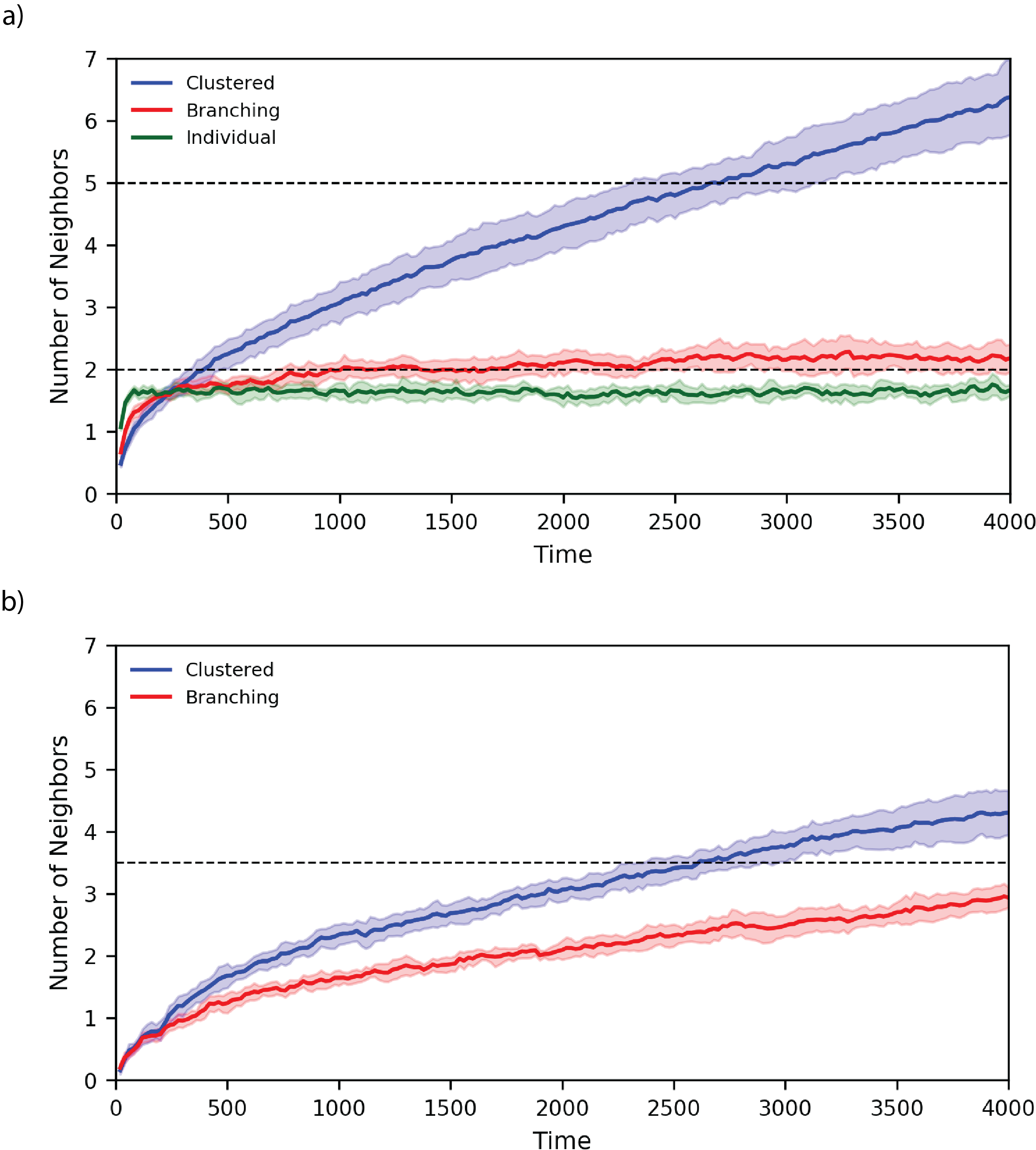}
\caption{\textbf{Number of neighbors over time.} (a) Local order parameter over time for individual, branching, and clustered phases at fixed population size. (b) Local order parameter over time for branching and clustered phases in a proliferating population. Dashed lines indicate threshold values to distinguish between individual, branching and clustered phases. Mean (solid bold line) and standard deviation (shaded region) computed using $10$ random initializations for each simulation.}
\label{fig:SI_Fig12}\vspace{-0.2in}
\end{figure*}

\begin{figure*}
\centering
\includegraphics[width=9cm]{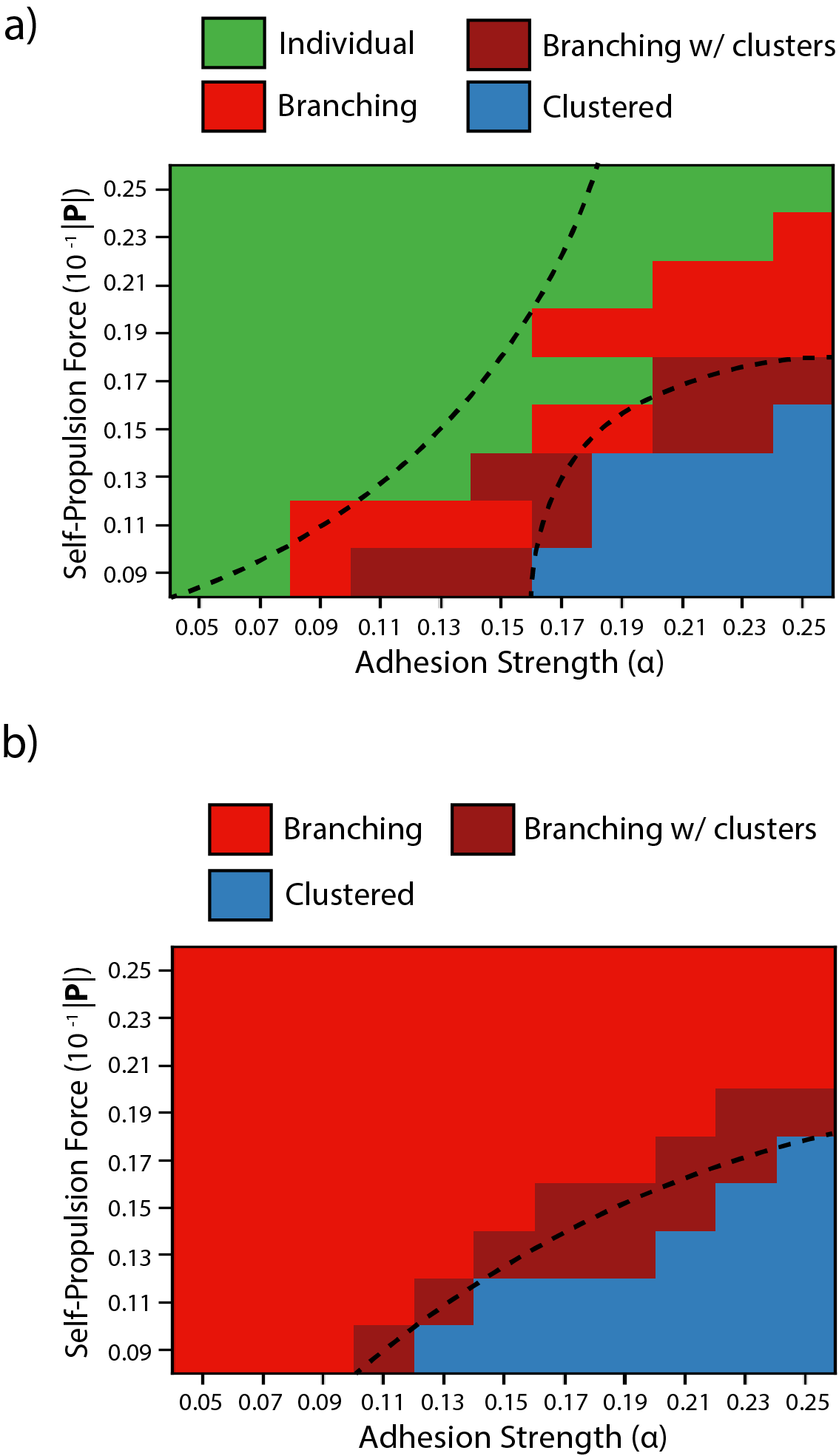}
\caption{\textbf{Phase diagram classified using dimension 0 homology (connected components).} (a) Simulations without proliferation are classified into $4$ distinct phases, individual, branching, branching with clusters, and clustered. (b) Simulations with proliferation are classified into $3$ distinct phases, branching, branching with clusters, and clustered. The dashed lines represent phase boundaries determined manually by inspecting particle positions. Classification results are biased by number of particles.}
\label{fig:SI_Fig_H0_phase}\vspace{-0.2in}
\end{figure*}

\begin{figure*}
\centering
\includegraphics[width=19cm]{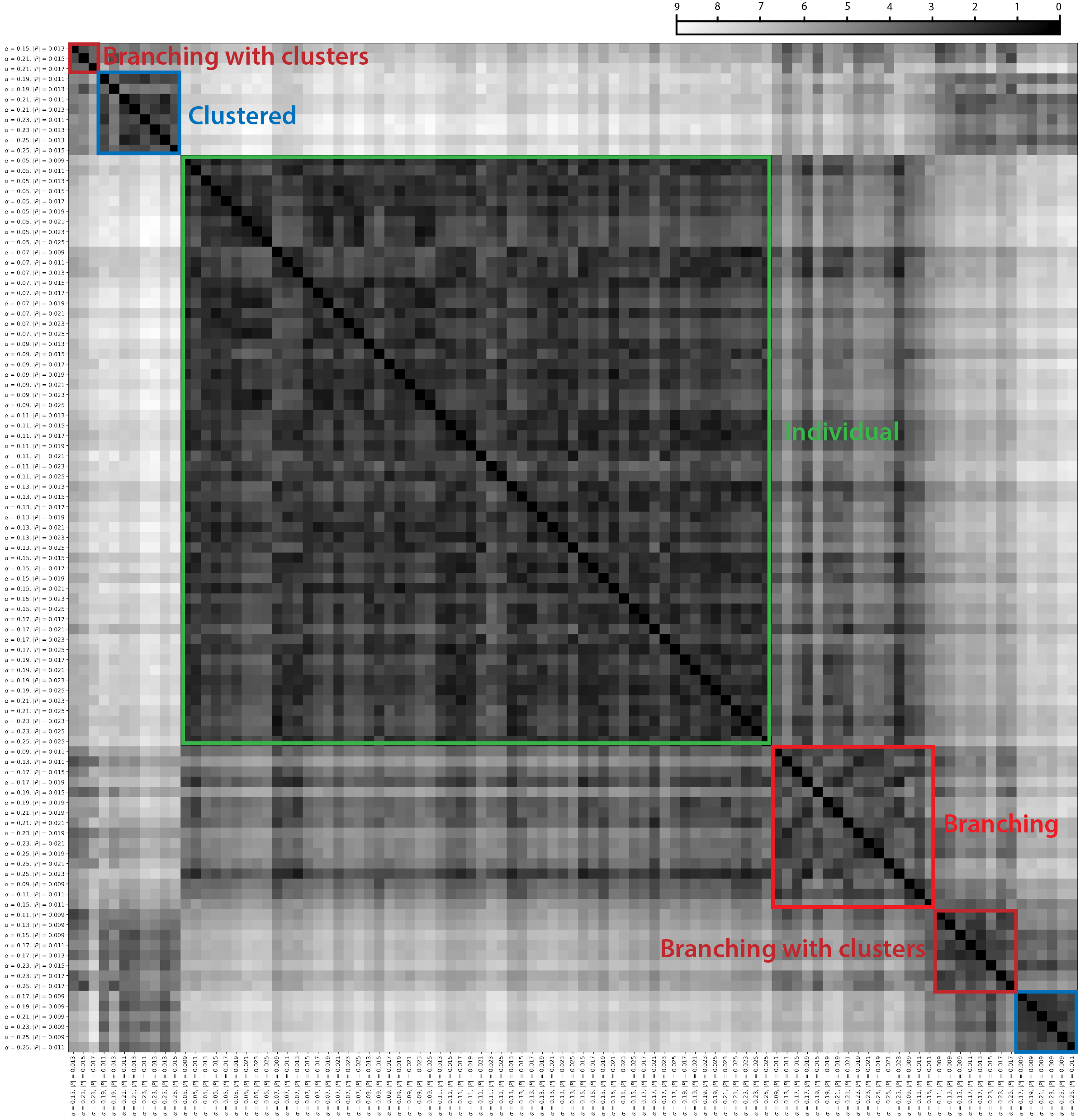}
\caption{\textbf{Pairwise Wasserstein distances computed using dimension 0 homology (connected components) between all simulations without proliferation, as well as varying adhesion and self-propulsion force.} Hierarchical clustering groups individual, branching, branching with clusters, and clustered corresponding to distinct phases along the diagonal. Note that branching with clusters phase exhibits similarity with both the branching and clustered phases.}
\label{fig:SI_Fig_noprolif_H0}\vspace{-0.2in}
\end{figure*}

\begin{figure*}
\centering
\includegraphics[width=19cm]{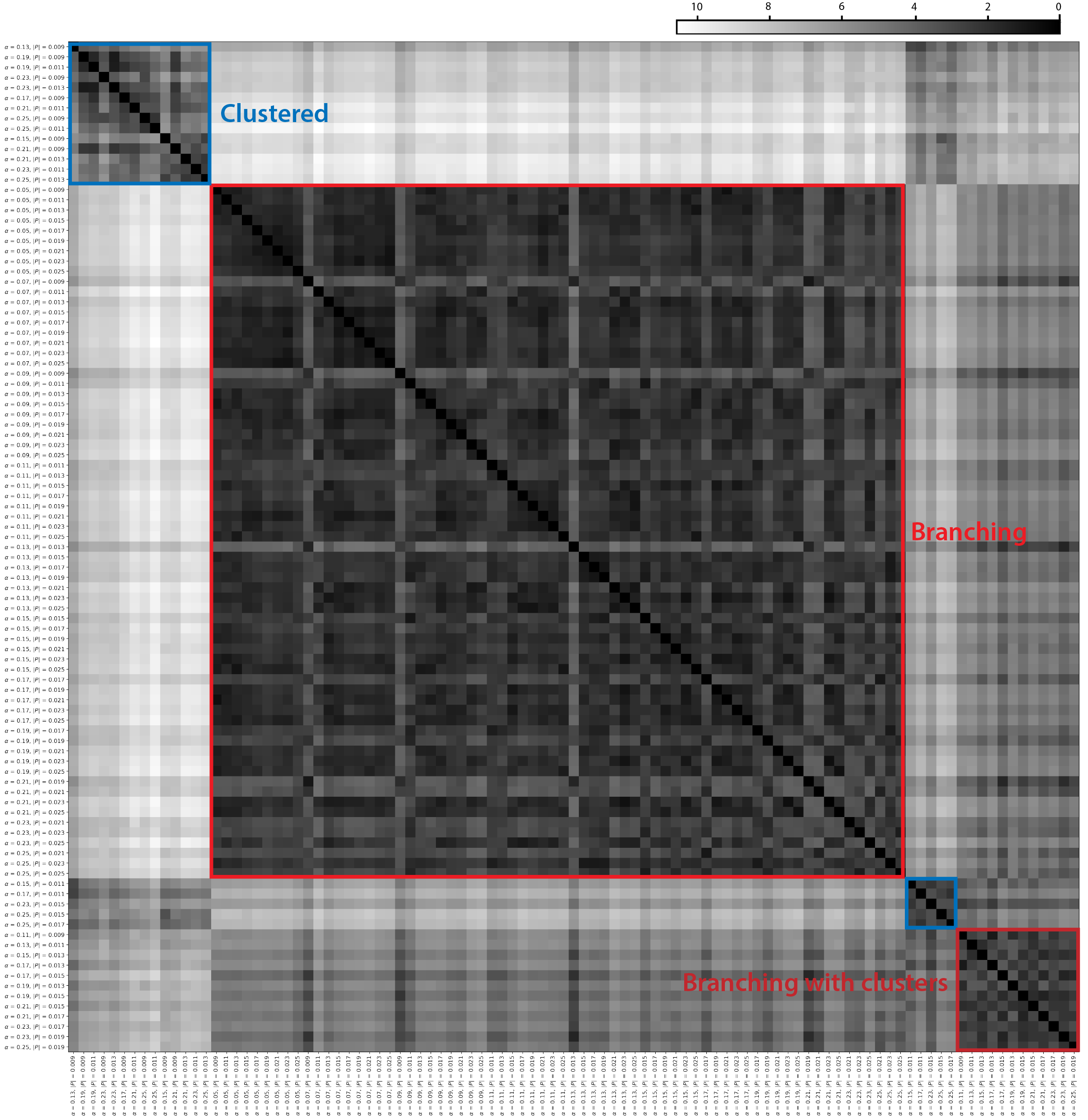}
\caption{\textbf{Pairwise Wasserstein distances computed using dimension 0 homology (connected components) between all simulations with proliferation, as well as varying adhesion and self-propulsion force.} Hierarchical clustering groups branching, branching with clusters, and clustered corresponding to distinct phases along the diagonal.}
\label{fig:SI_Fig_prolif_H0}\vspace{-0.2in}
\end{figure*}

To compare two persistence diagrams $X$ and $Y$, the notion of distance between these diagrams is defined using the Wasserstein metric as follows. First, a bijection, $s: X \to Y$, is defined by matching all off-diagonal points in $X$ with off-diagonal points in $Y$. Points close to the diagonal (corresponding to very short-lived and insignificant topological features) do not contribute to the distance between persistence diagrams. In case the two diagrams contain an unequal number of points, we also permit points to be matched to their projection on the diagonal, effectively ignoring them. Matching points across diagrams requires solution of an assignment problem, which is easier if the number of points in both diagrams are identical\cite{vidal_progressive_2020}. Therefore, in practice, projections of off-diagonal points to the diagonal are exchanged between persistence diagrams before matches are obtained (\textbf{Fig.~\ref{fig:SI_Fig2}, a-c}). The Wasserstein distance, $W(X, Y)$ is then defined as the infimum over all possible bijections, $s$:
\begin{equation}
W_{q,p}(X, Y) = \inf_{s: X \to Y} \left( \sum_{x \in X} \|x-s(x)\|^q_p \right)^{\frac{1}{q}}
\end{equation}
where for $p,q = 2$ we minimize the sum of squared Euclidean distances.

\clearpage

\section*{Note S2: System Parameterization based on Peclet Number and Attraction / Polarity}

\vspace{1cm}

The system is parameterized using two nondimensional variables corresponding to particle self-propulsion and adhesion. First, the Peclet number is defined in terms of the self-propulsion speed $P/\eta$, the particle radius $r_{eq}$, and the characteristic reorientation time $\tau_p$. The particle radius is given by the equilibrium separation distance where the interaction potential is a minimum, that is:

\begin{equation}
\label{eqn:F_mag}
|\mathbf{F}_{ij}| = - \Big|\frac{dU}{dr}\Big| = \alpha\Big|\frac{1}{4L_R}e^{-r/L_A}-\frac{1}{L_A}e^{-r/L_A}\Big| \quad 0 \le r \le r_{\text{max}}
\end{equation}

\begin{equation}
\label{eqn:r_eq}
|\mathbf{F}_{ij}| = 0 \implies r_{\text{eq}} = \frac{L_AL_R}{L_A-L_R}\ln\Big(\frac{L_A}{4L_R}\Big) \approx 1
\end{equation}

\noindent{Thus,}
\begin{equation}
\label{eqn:Pe}
\text{Pe} = \frac{\frac{P \tau_P}{\eta}}{r_{\text{eq}}} \approx 50 P
\end{equation}
where $\eta = 1$ and $\tau_P = 2500 \Delta t = 50$ is the characteristic time to repolarization. The Peclet number ranges between $0.4$ and $1.3$ for polarization values in our simulation.\\

Second a non-dimensional adhesion (scaled by self-propulsion), A, is defined that compares the relative strengths of adhesion and self-propulsion. The energetic cost $\Delta U$ of breaking a particle-particle bond is determined by moving the particle by a distance $\Delta r$ from $r_{eq}$ with $r_{max}$, beyond which the interaction energy is 0. Thus,

\begin{equation}
\label{eqn:U_eq}
U_{\text{eq}} = U(r_{\text{eq}}) = -\alpha \Big(\frac{4L_R}{L_A}\Big)^{\frac{L_R}{L_A-L_R}} + \frac{\alpha}{4}\Big(\frac{4L_R}{L_A}\Big)^{\frac{L_A}{L_A-L_R}} \approx -0.897\alpha
\end{equation}

\begin{equation}
\label{eqn:U_max}
U(r_{\text{max}}) = -\alpha e^{-r_{\text{max}}/L_A} + \frac{\alpha}{4}e^{-r_{\text{max}}/L_R} \approx -0.886\alpha
\end{equation}
We note that at any given time, the self-propulsion force $P$ is randomly oriented, thus we must average over all possible orientations $\theta$. When the self-propulsion force of the $j$th particle is oriented away from the $i$th particle, the effective force acting against the particle-particle bond is given by $P \cos(\theta)$ (i.e. $-\pi/2 < \theta < \pi/2$). However, when the self-propulsive force of the $j$th particle is oriented towards the $i$th particle, the effective force is 0. (i.e. $-\pi < \theta < -\pi/2$; $\pi/2 < \theta < \pi$). Thus,

\begin{equation}
\label{eqn:UoverP}
\text{A} = \frac{\Delta U}{\Delta r\int_{-\pi}^{\pi}P \cos(\theta) d\theta } = \frac{(U(r_{\text{max}}) - U(r_{\text{eq}})) \pi}{ P (r_{\text{max}}-r_{\text{eq}})} \approx \frac{0.011\alpha\pi}{P}
\end{equation}

\noindent $A$ ranges between $0.05$ and $2.1$ for $\alpha$ and $P$ values in our simulation.

\clearpage

\section*{Note S3: Runtime performance measurements for Wasserstein distance computation}

\noindent We computed pairwise Wasserstein distances between $121$ simulations corresponding to $11$ polarization values (linearly spaced between $0.005$ and $0.025$) and $11$ adhesion values (linearly spaced between $0.05$ and $0.25$). Computations corresponding to the two lowest polarization values ($P = 0.005, P = 0.007$) were later discarded from the phase diagram since they appeared to be kinetically trapped. Since the resulting distance matrix is symmetric, the Wasserstein distances were computed for the upper triangular part, including the diagonal, consisting of $121\times(121+1)/2 = 7381$ entries. Parallel computation was performed on $10$ CPUs (Intel Core i7-7800X, 3.5 GHz) with the i\textsuperscript{th} core responsible for computing entries in columns $12\times(i-1) + 1$ to $12\times(i)$. Therefore, core 1 computed entries in columns $1-12$ ($78$ values), core 2 computed entries in columns $13-24$ ($222$ values), and so on. Core 10 computed entries in columns $109-121$ ($1495$ values). In general, the i\textsuperscript{th} core was responsible for computing $144\times i - 66$ values, for values of i ranging from $1$ to $9$. Note that the 10\textsuperscript{th} core computed values in $13$ columns, whereas all other cores computed values in $12$ columns. While this is not the most efficient allocation of work (in an optimal configuration, work would be divided equally among the processors), it is easy to implement and produces results within a reasonable time frame (see Table \ref{tab:runtime_perf}).

\begin{table*}[h!]
\begin{center}
\begin{tabular}{c|c|c|c}
    Simulation & Barcode Dimension & \# Cores & Time (mins) \\
    \hline
    Without proliferation & 0 & 10 & 11 \\
    Without proliferation & 1 & 10 & 5 \\
    With proliferation & 0 & 10 & 36 \\
    With proliferation & 1 & 10 & 13 \\
\end{tabular}
\caption{\label{tab:runtime_perf}Time taken to compute pairwise Wasserstein distances.}
\end{center}
\end{table*}

\clearpage

\clearpage

\clearpage

\clearpage

\clearpage

\clearpage

\clearpage

\clearpage

\clearpage

\clearpage

\clearpage

\clearpage

\end{document}